\documentclass[11pt]{emulateapj}
\usepackage{graphicx}
\usepackage{amssymb,amsmath}
\usepackage{natbib}
\begin{document}
\topmargin 0.5in 

\title{Does size matter?  \\ The underlying intrinsic size distribution of radio sources and implications for unification by orientation}
\author{M.A.\ DiPompeo\altaffilmark{1}, J.C.\ Runnoe\altaffilmark{1}, A.D.\ Myers\altaffilmark{1}, T.A.\ Boroson\altaffilmark{2}}
\altaffiltext{1}{University of Wyoming, Dept. of Physics and Astronomy 3905, 1000 E. University, Laramie, WY 82071, USA}
\altaffiltext{2}{National Optical Astronomy Observatory, P.O. Box 26732, Tuscon, AZ 85726, USA}

\begin{abstract}
Unification by orientation is a ubiquitous concept in the study of active galactic nuclei.  A gold standard of the orientation paradigm is the hypothesis that radio galaxies and radio-loud quasars are intrinsically the same, but are observed over different ranges of viewing angles.  Historically, strong support for this model was provided by the projected sizes of radio structure in luminous radio galaxies, which were found to be significantly larger than those of quasars, as predicted due to simple geometric projection.  Recently, this test of the simplest prediction of orientation-based models has been revisited with larger samples that cover wider ranges of fundamental properties---and no clear difference in projected sizes of radio structure is found.  Cast solely in terms of viewing angle effects, these results provide convincing evidence that unification of these objects solely through orientation fails.  However, it is possible that conflicting results regarding the role orientation plays in our view of radio sources simply result from {\em insufficient sampling of their intrinsic size distribution}. We test this possibility using Monte-Carlo simulations constrained by real sample sizes and properties. We develop models for the real intrinsic size distribution of radio sources, simulate observations by randomly sampling intrinsic sizes and viewing angles, and analyze how likely each sample is to support or dispute unification by orientation.  \textit{We find that, while it is possible to reconcile conflicting results purely within a simple, orientation-based framework, it is very unlikely.} We analyze the effects that sample size, relative numbers of radio galaxies and quasars, the critical angle that separates the two subclasses, and the shape of the intrinsic size distribution have on this type of test.
\end{abstract}

\keywords{galaxies: jets, (galaxies:) quasars: general}

\section{INTRODUCTION}
Unification by orientation is the simplest picture historically invoked to explain the difference between different classes of Active Galactic Nuclei \citep[AGN; e.g.][]{Orr82,Ant85,Net85,Net87} and in particular, between luminous radio galaxies and radio-loud quasars \citep[e.g.][]{Sch79}.  In this scheme, radio galaxies (RGs) of the FRII type \citep{Fanaroff74} and radio-loud quasars (RLQSOs) are drawn from the same parent population, and each is only seen from a particular range of viewing angles with respect to the radio jet axis.  From along this axis (``pole-on'', viewing angle $\theta=0\arcdeg$) to some limiting angle $\theta_c$, one can see directly to the active nucleus and the optical properties will be consistent with radio-loud quasars.  Beyond $\theta_c$, to a completely ``edge-on'' ($\theta=90\arcdeg$) perspective, one will see optical properties of a radio galaxy. The limiting angle is assumed to be determined by the opening angle of an obscuring structure, such as a dusty torus, that blocks our view into the nuclear regions of RGs \citep[e.g.][]{Antonucci1993,Urry1995}.

Barthel (1989) published the first study constraining $\theta_c$ ($\sim45^{\circ}$) and seemingly confirming the simplest prediction of orientation unification; that the projected sizes of RLQSOs in radio images should be smaller than those of RGs, due purely to geometric effects. Although the idea that orientation could explain all properties of AGN remained somewhat controversial \citep[e.g.][]{Bor92}, unification through orientation has become quite popular for other types of objects beyond RGs and RLQSOs \citep[e.g.][]{Elv00}. The nature of this controversy remains  rich enough to have recently expanded to other tests over a wider parameter space, such as environmental measures over broad redshift ranges.  If orientation sufficiently explains AGN type, then the environments of different AGN should be identical throughout cosmic history. \citet{Wylezalek2013} demonstrated that RGs and RLQSOs occupy similar environments, although obscured AGN, which are often discussed within the ``dusty torus'' orientation framework, may have significantly different environments \citep{Hick11}.

Barthel's (1989) original test remains fertile ground, and newer, larger samples covering both a wider range in redshift and radio luminosity are now being used to revisit the simple projected size tests conducted over 20 years ago.  Both Singal et al.\ (2013) and Boroson et al.\ (2013, in prep) find results that contradict those of Barthel (1989), with independent samples that show no difference in projected radio-source size distributions (see \S 2 for more discussion of these results).  Therefore, they argue, factors other than orientation are necessary to differentiate between RGs and RLQSOs.

There is another possibility that can reconcile the seemingly contradictory size distribution results and keep the orientation by unification picture intact.  In the simplest case, all objects have a single intrinsic size.  Thus, if a viewing angle difference is present it will be manifested as a projected size difference in 100\% of experiments.  But, this single-size picture is highly idealized---so, assuming, instead, that samples of RGs and RLQSOs are drawn randomly  from some underlying {\em distribution} of intrinsic sizes (which is the same for both types of object), is it possible to produce both sets of results?  While the effect of viewing angle on the projected sizes of objects is the main focus of all previous work on this subject, \textit{what is the effect of the underlying intrinsic size distribution}?

The purpose of this paper is to explore these effects via Monte-Carlo modeling of randomly oriented objects generated from a reasonable intrinsic size distribution, using the properties of the samples in the three works mentioned above as constraints.  Our main goal is to explore statistically, given an intrinsic size distribution and assuming that unification by orientation is correct, how likely one is to conclude that orientation drives apparent radio-source sizes.  We also explore the effects of varying parameters such as the underlying intrinsic size distribution, the size of the sample, the relative number of RGs to RLQSOs in the sample, and the role of different values of $\theta_c$.  Finally, we attempt to determine the ideal sample that is needed to reliably test the paradigm of unification through orientation for RGs and RLQSOs.

We use a cosmology where $H_0 = 71$ km s$^{-1}$ Mpc$^{-1}$, $\Omega_M=0.27$, and $\Omega_{\Lambda} =0.73$ for all calculated parameters (Komatsu et al.\ 2011), but we note the use of even more recent cosmologies will make no difference in our results.  The notation $\langle$RG$\rangle$ and $\langle$RLQ$\rangle$ will be used throughout to indicate the median projected sizes of RGs and RLQSOs, respectively, in a sample or simulation.

\section{REAL SAMPLES}
We utilize three real samples that have been used for this kind of projected size test; those of Barthel (1989; B89), Singal et al.\ (2013; S13), and Boroson et al.\ (2013, in prep; B13; see also Boroson 2011).  We briefly summarize these below, but refer the reader to the original references for full details.  Some properties of the samples are summarized in Table~\ref{samples}.  See also the appendix for an analysis of the significance of the results in these works when median statistics are taken into account.

We acknowledge that there are other complications that could arise in projected size tests, beyond the effects of viewing angle and the distribution of intrinsic sizes.  There may be issues with how source sizes are measured, bending and rotation of radio jets, and sample selection biases.  Additionally, it is well known that there are two subclasses of radio galaxy--- the high-excitation and low-excitation radio-galaxies (HERGs/LERGs, respectively; Hine \& Longair 1979, Laing et al.\ 1983).  It is unlikely that LERGs participate in orientation unification schemes, as they probably don't harbor an AGN.  Singal et al.\ (2013) discuss this point explicitly.  However, none of the samples discussed here use HERGs explicitly.  As our goal here is to test whether a distribution of intrinsic sizes can reconcile already published results, we will not make any additional cuts, nor any re-measurements of sizes of the original sources.

\subsection{The B89 Sample}
This sample is drawn from the 3CRR catalog \citep{Laing83}, which is complete to $\sim$10.9 Jy at 178 MHz.  The high flux limit automatically restricts the sample to luminous radio sources at these redshifts (though not strictly FRII sources).   It includes all of the quasars (17) and radio galaxies (33) in the catalog with redshifts between $0.5 < z < 1.0$.  However, we note that there are two RGs in the sample with updated redshifts that place them outside of this range:  3C318 moves from $z=0.752$ to $z=1.574$ \citep{Willott2000}, and 3C325 moves from $z=0.86$ to $z=1.135$ \citep{Grimes2005}.
We include these here with the updated redshifts in order to keep the relative numbers the same as the original test, but excluding them (or including them with their original $z$ values) does not  significantly affect our results.

The results from B89 are consistent with an orientation-based scheme.  The projected size distribution of RGs is shifted toward larger values when compared to the QSOs, with the median RG size 2.2 times larger than the median RLQSO size ($\langle {\rm RG} \rangle / \langle {\rm RLQ} \rangle =2.2 $).  This is roughly consistent with the $\sim$45\arcdeg\ viewing angle separating the two classes derived from the relative number of objects in each class.

\subsection{The B13 Sample}
This sample is selected by matching radio data from 3 radio surveys with the Sloan Digital Sky Survey (SDSS). The radio surveys utilized are the Westerbork Northern Sky Survey (WENSS; Rengelink et al.\ 1997), Faint Images of the Radio Sky at Twenty-Centimeters (FIRST; Becker et al.\ 1995), and the NRAO VLA Sky Survey (NVSS; Condon et al.\ 1998). The low frequency observations of WENSS are used to select a sample unbiased in orientation, as these frequencies are not sensitive to beamed radio core emission. A match between the optical position and a radio core was not required, and a substantial fraction shows lobes but no core.  An explicit cut in radio luminosity ($\log L$(325 MHz) $> 26.5$ W Hz$^{-1}$) excludes low-luminosity radio sources. This threshold is well above the flux limit of WENSS within the redshift range ($0.1 < z < 0.5$) of the sample.  Only objects targeted with SDSS spectra as part of complete samples are included, and comparisons with other radio samples indicate that the only objects that are missing are a small number of optically faint radio galaxies.  Essentially all of the objects have steep radio spectra and FR II radio morphology.  The sample includes 51 RGs and 35 RLQSOs, which can be distinguished by the SDSS spectroscopy.  The angular sizes were measured from the FIRST images, which have sufficient resolution to easily resolve them at these redshifts.

The findings using this sample contradict the original test by B89. RLQSOs are found to be larger than RGs by a median factor of 1.6 ($\langle {\rm RG} \rangle / \langle {\rm RLQ} \rangle=0.6$), which is used to argue that RLQSOs cannot be the same objects as RGs seen from smaller viewing angles.

\subsection{The S13 Sample}
This sample is drawn from the Molonglo Reference Catalog (MRC; Kapahi et al.\ 1998).  The catalog is also flux limited at low-frequency, but deeper than 3CRR (complete to $\sim$0.95 Jy at 408 MHz).  However, the depth is shallow enough that only luminous radio sources of the type considered for unification are included.  The redshift range is larger than that of the B89 sample, with $0 < z < 3.2$.  The full sample includes 379 RGs and 87 RLQSOs, though various subsamples are also studied (different redshift ranges, radio luminosities, excluding compact steep spectrum (CSS) sources, etc.).  We will only consider the full sample here as results are consistent amongst the various subsets considered by S13.

The S13 results contradict B89 as well, as no difference is seen in the projected size distributions, even when limited to the same redshift range.  Again, any difference suggested by the data indicates that the RLQSOs are \textit{larger} by a median factor of 1.3 ($\langle {\rm RG} \rangle / \langle {\rm RLQ} \rangle=0.8$), roughly in agreement with B13.

\begin{deluxetable*}{lcccccccc}
\centering
 \tabletypesize{\footnotesize}
 \tablecaption{Basic properties of the real samples used\label{samples}}
    \tablehead{
\colhead{Sample}  &  \colhead{$n_{tot}$}  &  \colhead{$n_{RG}$}  &  \colhead{$n_{RLQ}$}  &  \colhead{$f_{RG}$}  & \colhead{$f_{RLQSO}$}  &  \colhead{$z$ range} & \colhead{$\langle RG \rangle / \langle RLQ \rangle$} &  \colhead{Reference}
}
\startdata
B89  &    50    &    33    &    17    &    0.66    &    0.34    &    $0.5$-$1.5$\tablenotemark{a}     & 2.2   &   Barthel et al.\ (1989) \\
B13  &   86     &   51     &    35     &   0.59    &    0.41    &   $0.1$-$0.5$                                      & 0.6   &   Boroson et al.\ (in prep)\\
S13  &  466    &  379   &    87     &   0.81    &    0.19    &    $0.1$-$3.2$                                     & 0.8   &  Singal et al.\ (2013) 
\enddata
\tablenotetext{a}{The original reference quotes a redshift range of $0.5$-$1.0$; however, two of the RGs have updated values of $z$ extending the range to $\sim$1.5.  Including these with the new values, keeping the old values, or excluding them altogether has little effect on our results.}
\end{deluxetable*}

\section{INTRINSIC SIZE DISTRIBUTIONS}
\subsection{Modeling the samples used}
In order to build a distribution of intrinsic sizes to draw from, we start with the data from B89, S13 and B13.  We use the apparent angular sizes and redshifts of each object, and assign each a random viewing angle consistent with a purely geometrically defined distribution (i.e. an edge-on orientation is much more likely than a pole-on one; $N(\theta) = \sin \theta$).  An intrinsic size is calculated for each object, and the process is repeated $10^3$ times.  The resulting distribution of sizes is shown for each sample in the first three panels of Figure~\ref{intrinsicmodels}, in 50 kpc bins.  This method essentially provides the ``widest'' possible distribution of sizes, as the large number of iterations allows each object to be assigned a wide range of orientations.  Imposing a viewing angle restriction on the RG and RLQSO subsamples (i.e.\ only allowing RGs to be seen from 90 to $\theta_c$ degrees and RLQSOs from $\theta_c$ to 0 degrees) does not change the resulting shape substantially.  It simply changes which objects populate which part of the distribution.  This is due to the fact that the apparent angular size distributions of the two subsets are similar.

We next fit functions to these generated distributions.  The forms of the fits are determined purely empirically, and we assign them no physical significance.  We find that two functions are needed in different regions to adequately fit the distributions.  In natural logarithmic space, below 500 kpc a power law is used, and above 500 kpc a linear fit is used.  The actual size at which the break between functions occurs is determined by where the fits are closest in $N$ (not necessarily at 500 kpc).  A cut at a maximum size of 3 Mpc is also imposed, as sources larger than this are extremely rare; however, we find that where this cut is placed has very little effect on our results.  The final exponentials are then normalized by the largest value of $N$.  The fit parameters to this ``wide'' model are given in the middle portion of Table~\ref{isizemodels}, and the fits are shown as dashed lines in Figure~\ref{intrinsicmodels}.

As mentioned above, these fits give an upper limit to the size distributions, and may give an unrealistic number of large sources.  If we take 40 kpc as a split between large and small sources, then indeed these fits result in distributions where 80--90\% of the sources are considered large.  A more conservative large/small ratio for luminous radio sources is 40--50\% (Bridle \& Perley 1984).  Therefore, we also determine, purely empirically and using the same functional forms, fit parameters for each sample that provide, on average, this ratio when the real sample numbers are randomly drawn from the distributions.  These ``narrow'' fit parameters are given in the top portion of Table~\ref{isizemodels}, and shown as dotted lines in Figure~\ref{intrinsicmodels}.

Finally, the fit parameters for the three samples are averaged together (separately for the wide and narrow distributions).  These mean fits are used in the simulations to define the underlying intrinsic size distribution, and are shown in bold in Table~\ref{isizemodels}.

\subsection{Modeling a pure FRII sample}
In order to completely separate the process of developing an intrinsic size distribution from the samples we are trying to simulate, we develop a third distribution from an independent sample.  Mullin et al.\ (2008; M08) present the properties of a sample of strictly FRII type sources (including RGs and QSOs).  As these are the types of objects for which this kind of projected size difference test is intended, it is an ideal sample to utilize in building an intrinsic size distribution.  While the other samples above will contain many powerful FRII type sources, they may not be restricted to \textit{only} these objects.

\begin{figure}
\vspace{-0.2cm}
\begin{minipage}{8.5cm}
\centering
\includegraphics[width=8.5cm]{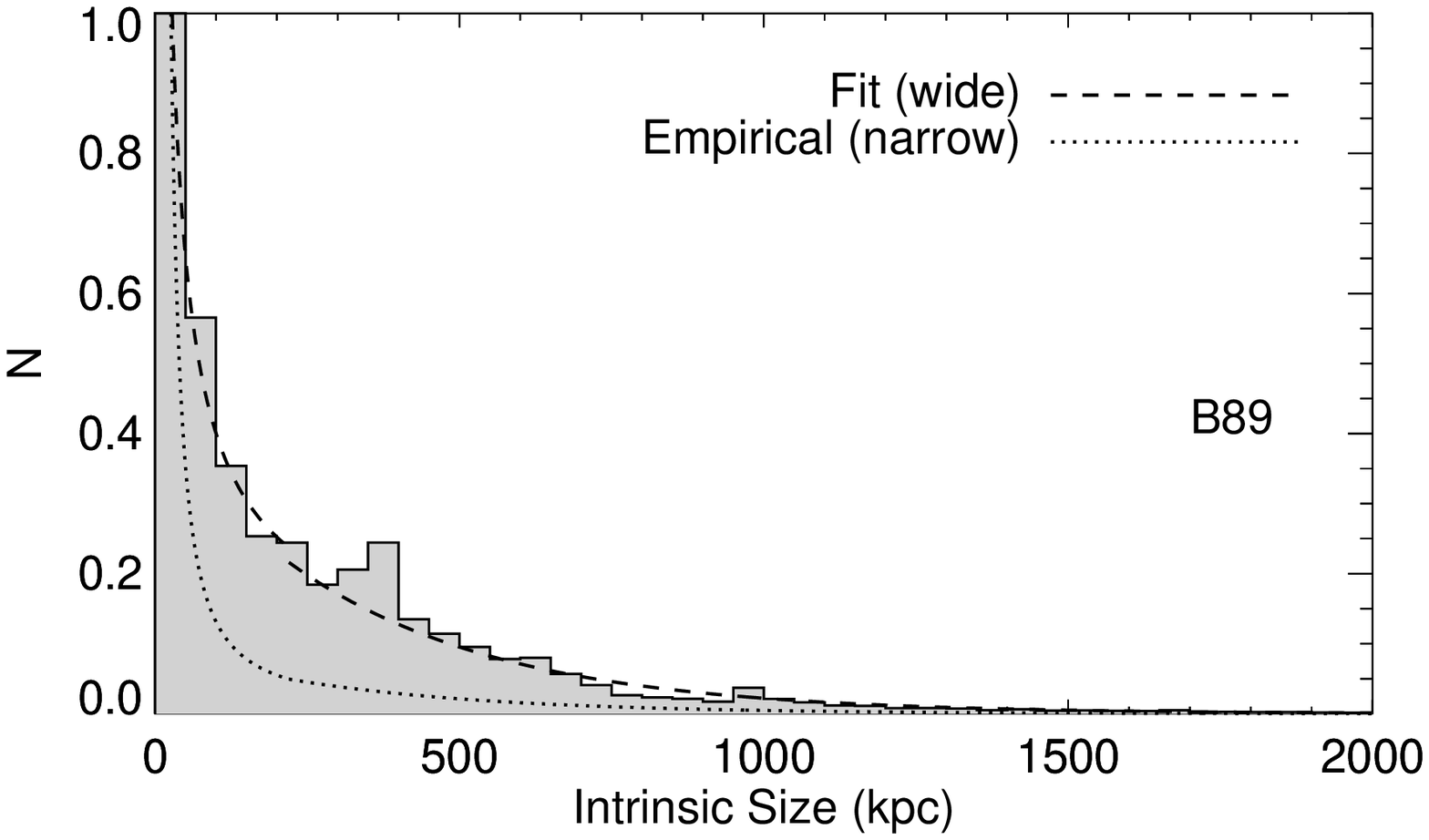}
\end{minipage}
\vspace{-0.4cm}

\begin{minipage}{8.5cm}
\centering
\includegraphics[width=8.5cm]{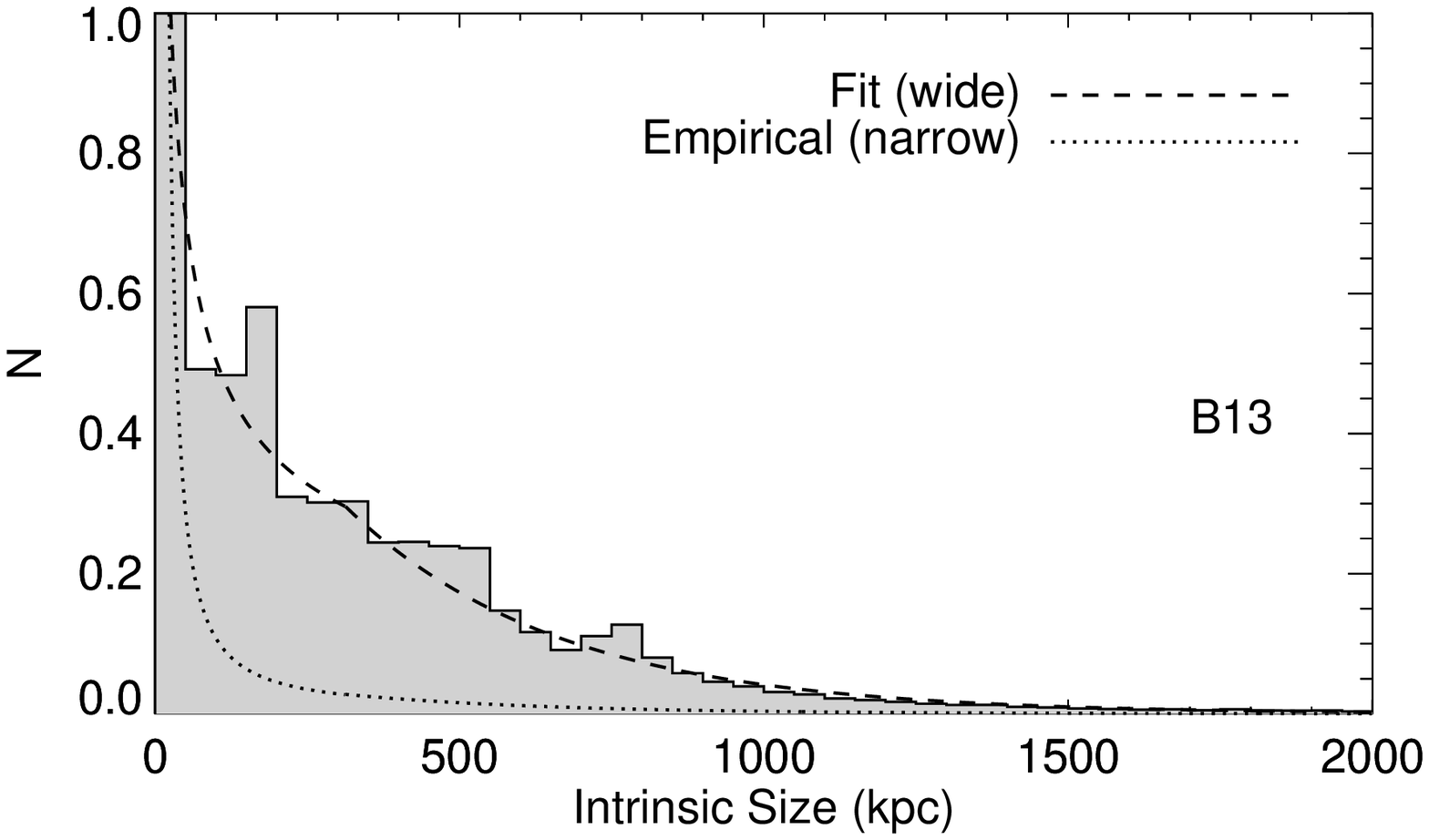}
\end{minipage}
\vspace{-0.4cm}

\begin{minipage}{8.5cm}
\centering
\includegraphics[width=8.5cm]{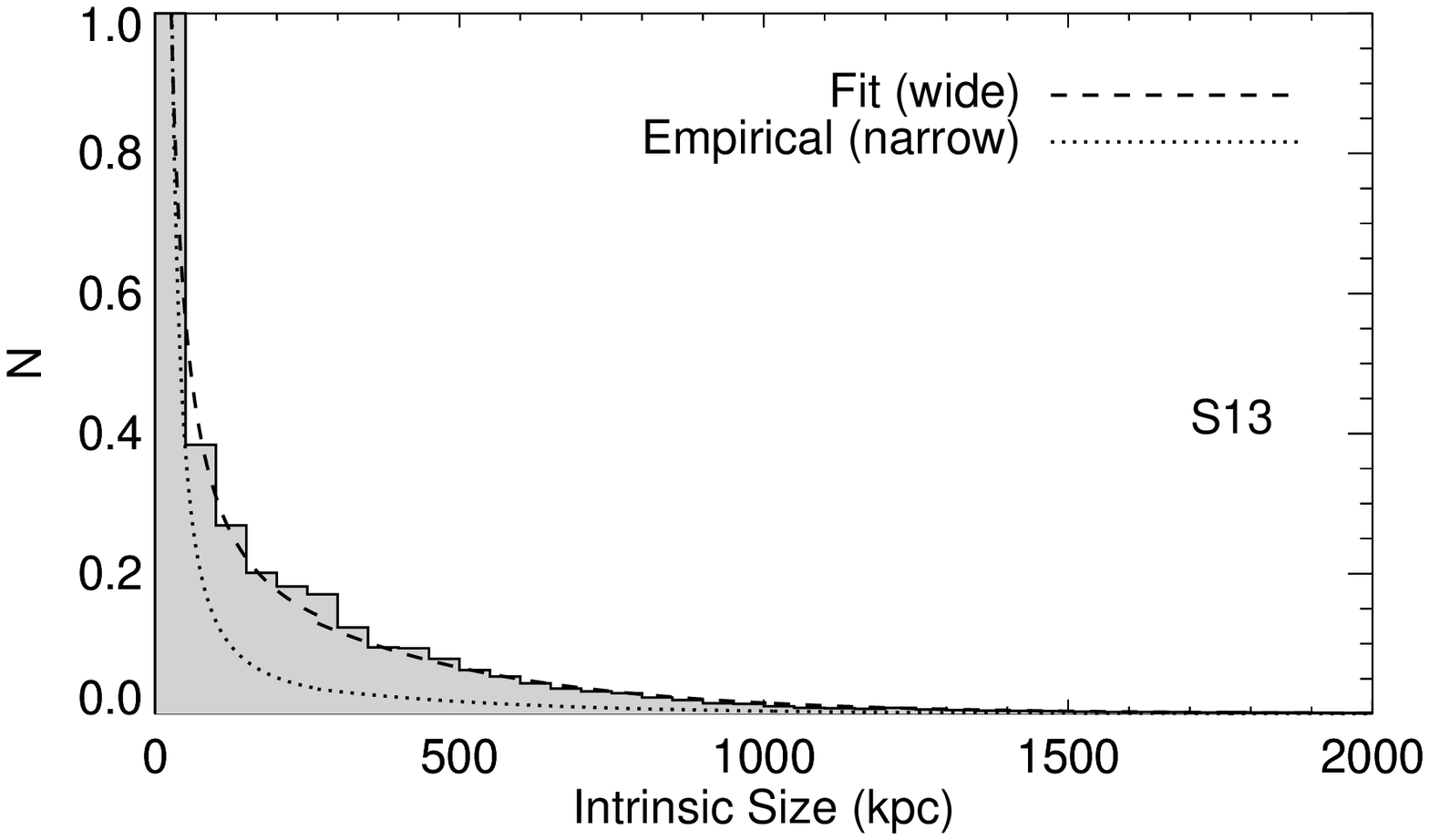}
\end{minipage}
\vspace{-0.4cm}

\begin{minipage}{8.5cm}
\centering
\includegraphics[width=8.5cm]{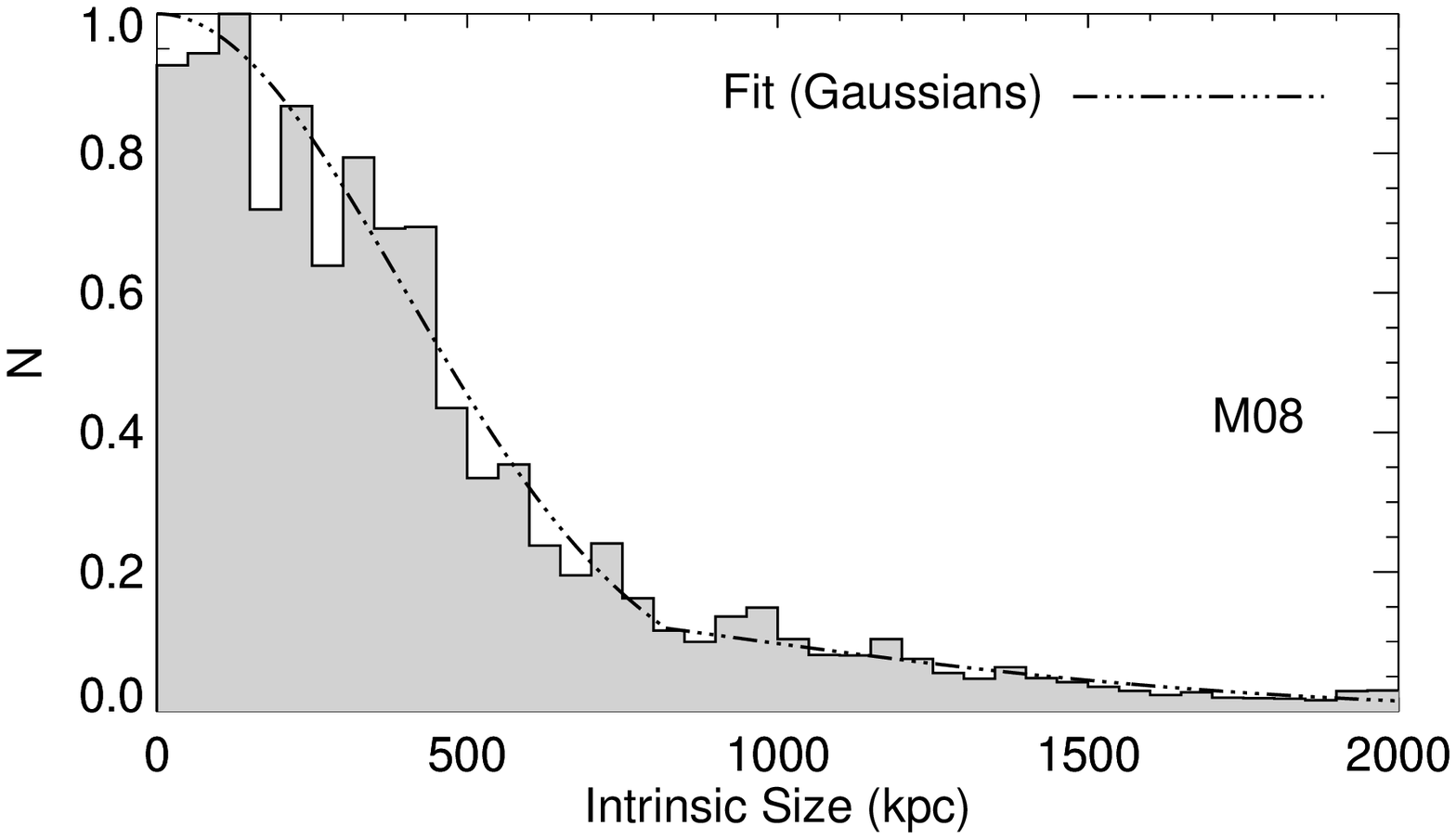}
\end{minipage}
\vspace{-0.2cm}
\caption{Intrinsic size distributions generated by modeling.  Top three panels: Each object in the B89, S13, and B13 samples is assigned a random viewing angle consistent with geometric effects, and the distance and apparent angular size are used to determine an intrinsic size. This process is repeated $10^3$ times for each sample, and results in the shaded distributions.  The dashed lines are fits consisting of a power law (at smaller sizes) and a straight line (at larger sizes) in natural-logarithmic space, whose parameters are given in Table~\ref{isizemodels}.  This is taken as the widest possible version of the intrinsic size distribution.  The dotted line is a narrower distribution, defined such that each sample has $\sim$50\% of objects above and below 40 kpc.  Bottom panel: the same method applied to a pure FRII source sample---this is best fit by two Gaussian components, shown as dot-dashed lines.\label{intrinsicmodels}}
\end{figure}

We apply the same method as in the last subsection to turn the apparent linear sizes presented in M08 into intrinsic sizes, by assigning each object a random viewing angle and repeating the process $10^3$ times to build a distribution.  The result is shown in the bottom panel of Figure~\ref{intrinsicmodels}.  No longer does the distribution have the same functional form as the previous three samples, as there are quite a few more intermediate (100--500 kpc) sized sources, as well as quite a few more large ($>1500$ kpc) sources.  This is reflective of the restriction to only FRII  objects, which are expected to be large.  We find that two truncated Gaussians accurately describe this distribution, one below $\sim$800 kpc, and another above.  The parameters of these Gaussians are given at the bottom of Table~\ref{isizemodels} and shown by dot-dashed lines in Figure~\ref{intrinsicmodels}.

\begin{deluxetable*}{lcccccccc}
\centering
 \tabletypesize{\footnotesize}
 \tablecaption{Intrinsic Size Distribution Models\label{isizemodels}}
    \tablehead{
   \colhead{}               & \colhead{}                & \multicolumn{3}{c}{$N=K_{1}e^{c s^{\alpha}}$}  & \colhead{} & \multicolumn{3}{c}{$N=K_{2}e^{m s + b}$} \\
                                                                                                  \cline{3-5}                                                                 \cline{7-9}       \\   
   \colhead{Sample} & \colhead{Break (kpc)} & \colhead{$K_1$} & \colhead{$c$} & \colhead{$\alpha$} & \colhead{} & \colhead{$K_2$} & \colhead{$m$} & \colhead{$b$} 
   }   
   \startdata
          &  &  \multicolumn{7}{c}{Empirically determined (``narrow'')}\\
                   \cline{3-9}   \\
   B89     &  220 &  $7.9 \times 10^{-5}$ & 17.00  & $-$0.18  &  &  $7.9 \times 10^{-5}$  &  $-$0.003  &  7.10  \\
   B13     &  305 &  $6.5 \times 10^{-5}$ &  17.00  & $-$0.18 &  &   $6.5 \times 10^{-5}$ &  $-$0.003  &  7.00  \\
   S13     &  270 &  $9.7 \times 10^{-6}$ & 19.00  & $-$0.15  &  &   $9.7 \times 10^{-6}$ &  $-$0.003  &  9.00  \\
\textbf{Mean}    &   \textbf{270} &  $\mathbf{5.1 \times 10^{-5}}$ &  \textbf{17.67}  & $\mathbf{-0.17}$ &  &   $\mathbf{5.1 \times 10^{-5}}$ &  $\mathbf{-0.003}$  &  \textbf{7.70}  \\
           &   &  \multicolumn{7}{c}{Fit to model data (``wide'')}  \\
               \cline{3-9}   \\
   B89     &  220 &  $7.9 \times 10^{-5}$ & 12.36  & $-$0.08  &  &  $7.9 \times 10^{-5}$  & $-$0.003  &  8.54  \\
   B13     &  305 &  $6.5 \times 10^{-5}$ &  11.49 & $-$0.05  &  &   $6.5 \times 10^{-5}$ &  $-$0.003  &  9.31  \\
   S13     &  270 &  $9.7 \times 10^{-6}$ & 15.06  &  $-$0.08 &  &   $9.7 \times 10^{-6}$ &  $-$0.003  &  10.27  \\
\textbf{Mean}     &  \textbf{273} &  $\mathbf{5.1 \times 10^{-5}}$ &  \textbf{12.97} & $\mathbf{-0.07}$  &  &   $\mathbf{5.1 \times 10^{-5}}$ &  $\mathbf{-0.003}$  &  \textbf{9.38}   \\
 \hline 
          & &  \multicolumn{7}{c}{Modeled Gaussians ($N=Ae^{-(s-\mu)/2\sigma^2}$)} \\
          \cline{3-9}  \\
          &  &  \multicolumn{3}{c}{Below break} & & \multicolumn{3}{c}{Above break} \\
          &               &  A    &    $\mu$    &   $\sigma$    & &   A       &   $\mu$    &   $\sigma$    \\   
               \cline{3-5}                                      \cline{7-9} \\
  \textbf{M08}   &   \textbf{820}  &   \textbf{1}   &      \textbf{0}          &     \textbf{397.8}         & &  \textbf{0.18}  &       \textbf{0}         &     \textbf{896.8}           
    \enddata
 \tablecomments{Parameters describing the modeled intrinsic size distributions.  The distributions generated using the B89, S13, and B13 samples, in natural log space, are best described by a power law below some break (given in column (2)), and a straight line above the break.  Columns (3)-(5) give the parameters of the power-law fit, and columns (6)-(8) give the parameters of the linear fit.  The middle portion of the table gives these parameters using the real fits to the distributions, which describes the widest possible intrinsic size distribution for the samples.  The top portion gives the parameters empirically determined to keep the same functional form but force a roughly even split in the resulting distribution between large ($>40$ kpc) and small ($< 40$ kpc) sources.  The bottom portion shows the Gaussian fits above and below the break used to describe the intrinsic sizes generated with the FRII sample of M08.}
\end{deluxetable*}

\section{SIMULATIONS \& RESULTS}
\subsection{Simulating data samples}
\subsubsection{Methodology}
Our general approach is to perform simulated versions of the original B89, S13, and B13 tests while varying several parameters.  In all cases we use the real redshifts of objects in the samples, which are converted into angular size distances.  We assign each object an intrinsic size by drawing from one of the three (wide, narrow, or Gaussian) distributions discussed in the previous section, and a random viewing angle drawn from a distribution consistent with geometric projection effects.  These parameters are combined to calculate both the projected size on the sky in kpc and the apparent angular size in arcseconds for each object.  We calculate the ratio of the median RG projected size to median RLQSO projected size.  The distributions of projected sizes and angular sizes for the RGs and RLQSOs are also compared via a Kolmogorov-Smirnov (KS) test.  This process is repeated $10^5$ times.  The final use of these values will be described in detail in the following subsections.

Every set of $10^5$ tests is run drawing from the narrow, wide, and Gaussian distributions of intrinsic size separately.  We also run each set using various values\footnote{As a reminder, we constrain RGs to have $\theta_c < \theta < 90\arcdeg$, and RLQSOs to have $0 < \theta < \theta_c$} for $\theta_c$; 30$\arcdeg$, 45$\arcdeg$, and 60$\arcdeg$.  Additionally, we run the simulations assuming no viewing angle restriction on RGs or QSOs (i.e.\ unification by orientation is false).  These results will be labeled throughout as ``no restriction'' or ``no $\theta_c$.

Finally, we run each test using the original sample size parameters (total number of objects $n_{tot}$, number of RGs $n_{RG}$, number of RLQSOs $n_{RLQ}$), as well as resampling them to have an equal number of RGs and RLQSOs but keeping the same $n_{tot}$.  This resampling is conducted by simply selecting objects randomly from each subset.  The $z$ distributions in this random sampling are not forced to be matched, however we find that in most cases they are.  A new random sample is drawn for each of the $10^5$ iterations.  We note that, to estimate the value of $\theta_c$ in the orientation scenario, as B89 did, one must carefully select every RG and RLQSO available in a sample without bias---but to look at projected size distributions this is less critical, so long as selection biases do not affect the RG and RLQSO distributions differently.  We are interested in the latter question here.  So while it may not necessarily be possible (or reasonable) to force a sample to have an equal number of objects, we are merely interested in seeing how this ratio can affect the outcome.

All of our results are presented using the linear projected sizes, and not the apparent angular sizes.  In most cases, the results using these values are quite similar, except using the S13 sample where differences in the apparent sizes are more pronounced when using the angular sizes.  This is due to the redshift matching of the RGs and RLQSOs.  While statistically the S13 sample has well-matched distributions of redshifts for both RGs and RLQSOs, there is a notable overabundance of low-$z$ RGs, which they discuss in their paper.  If we neglect some of these objects, and force the $z$ distributions to match better by eye, this difference between using projected and angular sizes is diminished.

\subsubsection{Distribution of $\langle RG \rangle / \langle RLQ \rangle$}
\begin{figure}
\vspace{-0.2cm}
\begin{minipage}{8.5cm}
\centering
\includegraphics[width=8.5cm]{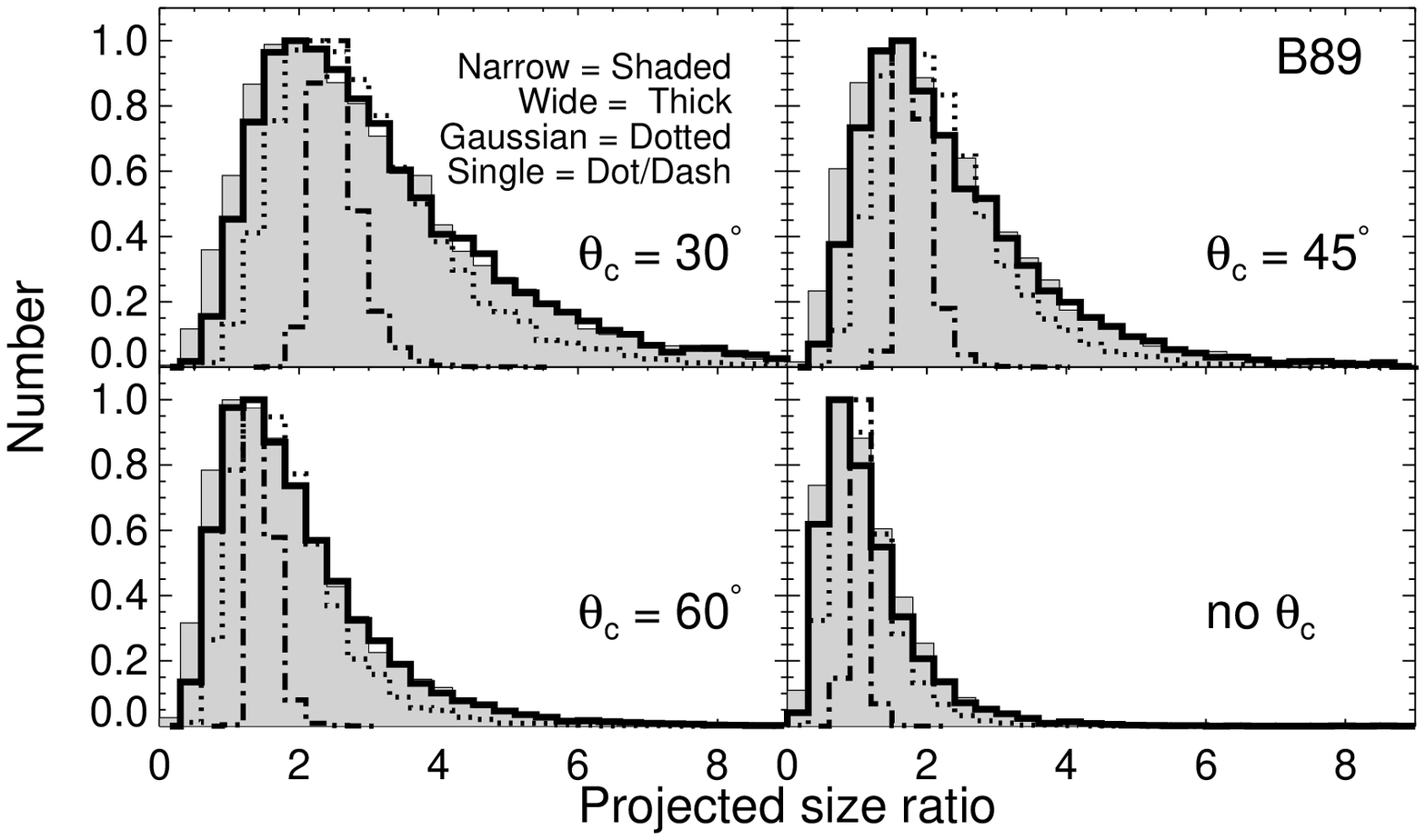}
\end{minipage}
\vspace{-0.4cm}

\begin{minipage}{8.5cm}
\centering
\includegraphics[width=8.5cm]{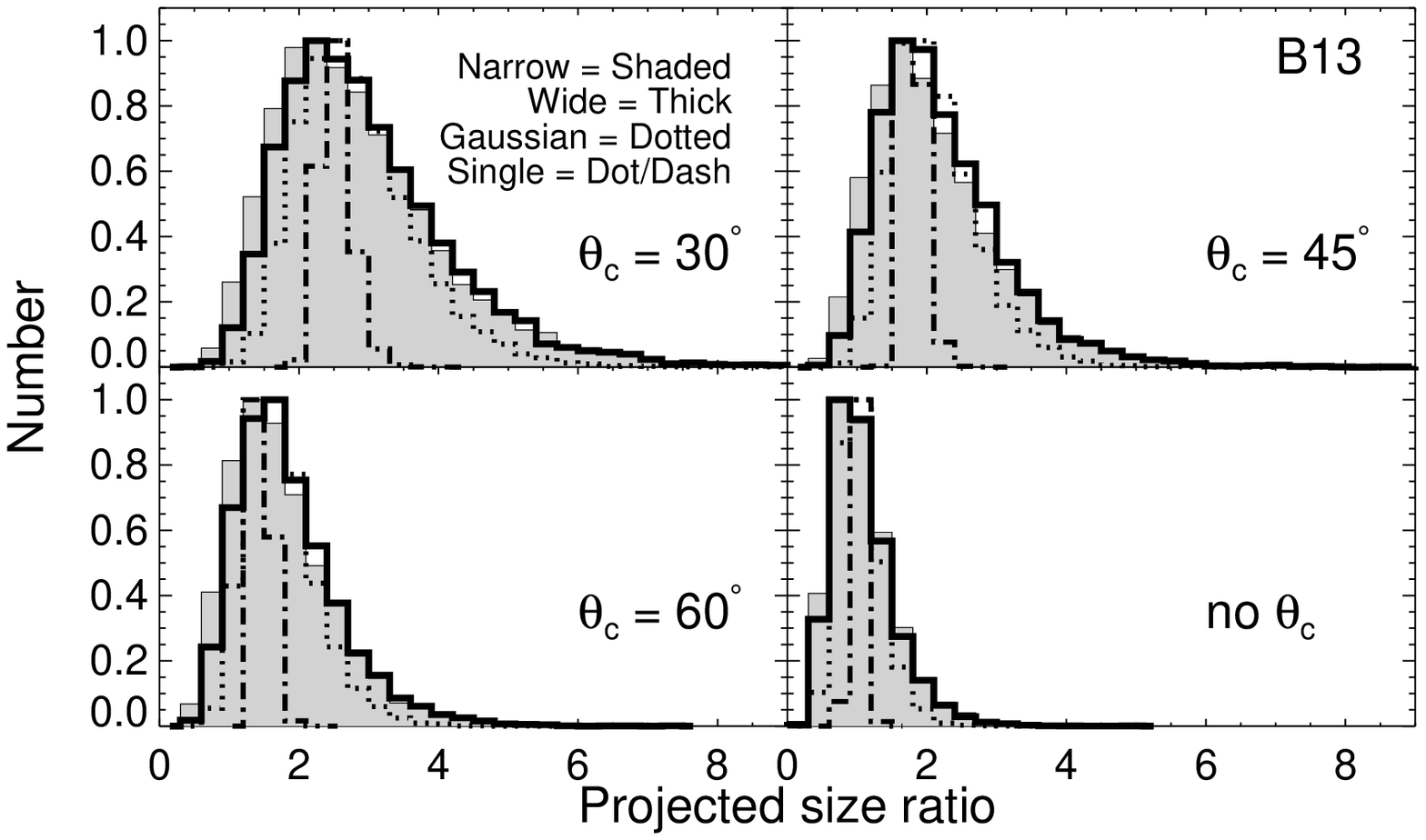}
\end{minipage}
\vspace{-0.4cm}

\begin{minipage}{8.5cm}
\centering
\includegraphics[width=8.5cm]{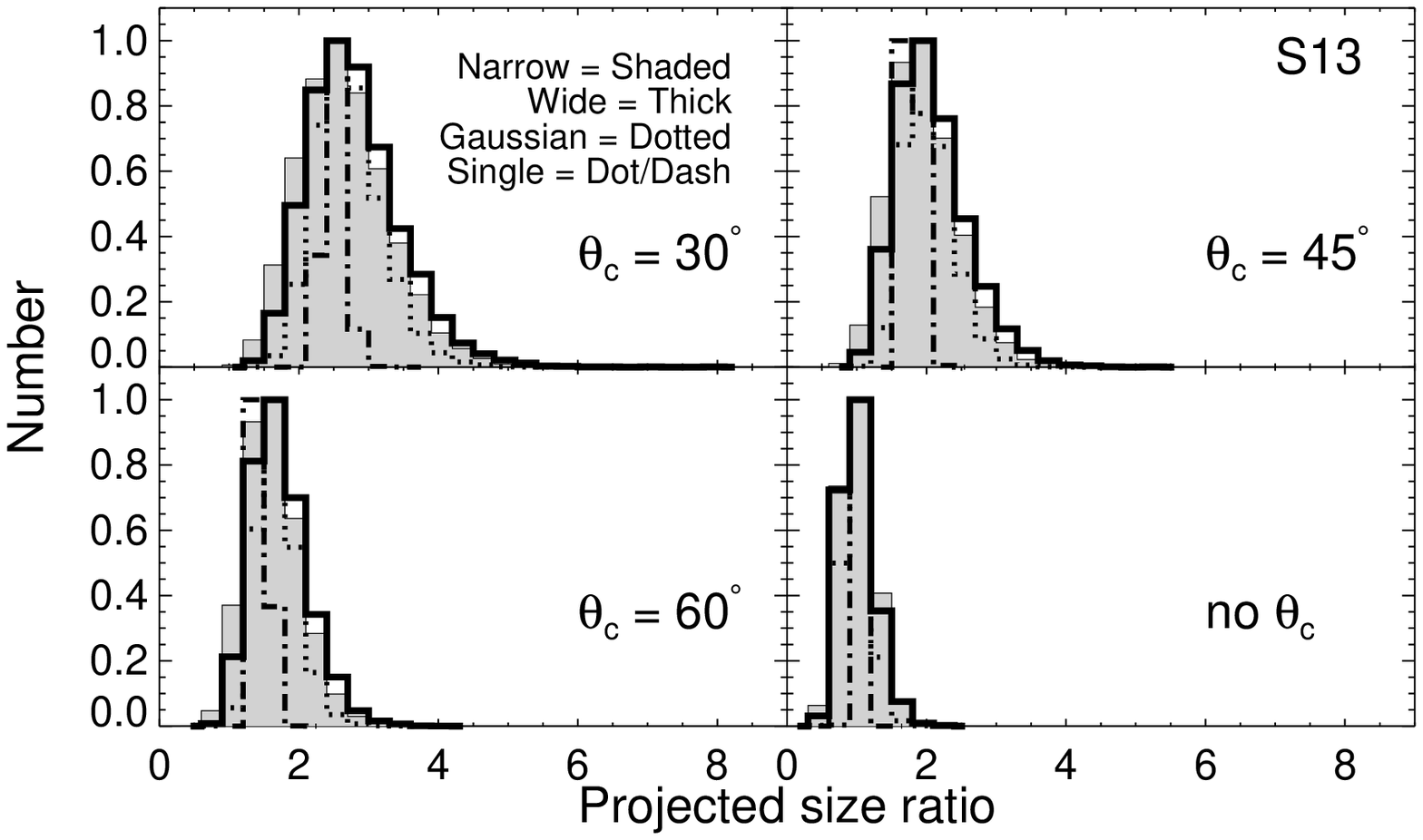}
\end{minipage}
\vspace{-0.2cm}
\caption{The distributions of median RG to RLQSO projected size for each sample, value of $\theta_c$, and intrinsic size distribution.  Also shown are the results for a single input intrinsic size, as opposed to a distribution with significant width.  We only show these distributions for the original sample sizes and $n_{RG}/n_{RLQ}$ values, as they are nearly identical for the resampled 1:1 ratio simulations.\label{meddists}}
\end{figure}

Most of our results are drawn from the distributions of $\langle {\rm RG} \rangle / \langle {\rm RLQ} \rangle$, which are shown in Figure~\ref{meddists} with the peaks normalized to one in order to help illustrate the subtle change in shape when varying the intrinsic size distribution.  These distributions are built by taking the ratio of the median RG projected size to the median RLQSO projected size in each of the $10^5$ iterations.  The histograms are converted into probability density functions (PDFs) by normalizing their area to 1, allowing us to simply integrate within various limits to determine the probability of a particular outcome.  The PDFs normalized by area are not shown explicitly here, as their shape is the same as the histograms in Figure~\ref{meddists} and the reader can get a sense of the probabilities from there.

We also show in Figure~\ref{meddists} the $\langle {\rm RG} \rangle / \langle {\rm RLQ} \rangle$ distributions when a single intrinsic value is used as opposed to a range.  Note that they have some width due to variation in the viewing angles, but they are quite narrow in comparison.

\begin{figure}
   \includegraphics[width=8.5cm]{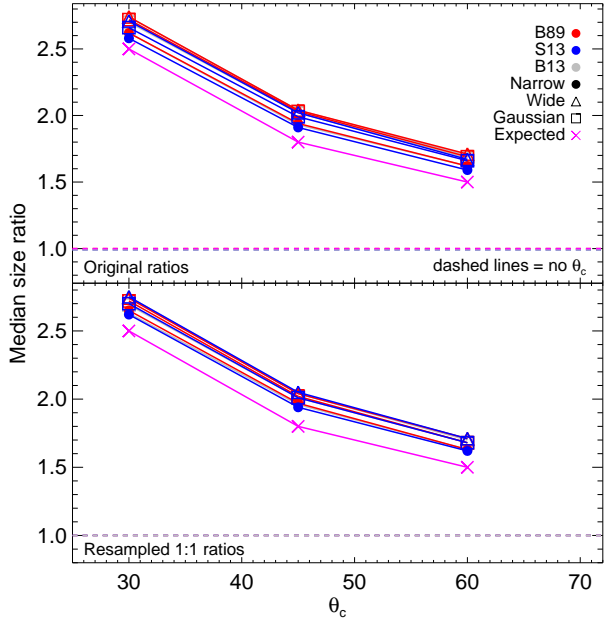}
  \caption{The median values of the median RG/RLQSO projected size ratios.  The top panel shows results keeping the original $n_{RG}/n_{RLQ}$ ratios for each sample, and the bottom shows the results keeping the same $n_{tot}$ but randomly resampling to an equal number of RGs and RLQSOs.  The ``expected'' values indicate what the mean projected size ratio would be in the simplest case where there is a single intrinsic size for every source.\label{medresults}}
\end{figure}

The medians of these $\langle {\rm RG} \rangle / \langle {\rm RLQ} \rangle$ distributions provide some insight into the effects of the intrinsic size distributions, as well as some checks that our simulations are performing as expected, and are shown in Figure~\ref{medresults}.  In the simplest scenario where all RGs and RLQSOs have a single intrinsic size, not a distribution, one can predict what the most probable median projected size ratio will be.   For a given $\theta_c$, one can directly find the median viewing angle ($\langle\theta \rangle$) to RGs or RLQSOs, using the fact that the distribution of possible viewing angles obeys the relationship $N(\theta)=\sin \theta$.  The expected ratio of median sizes is then $\sin (\langle \theta_{RG}\rangle) / \sin (\langle \theta_{RLQ}\rangle)$.  This predicts median projected size ratios of 2.5, 1.8, and 1.5, for $\theta_c= 30\arcdeg, 45\arcdeg$, and $60\arcdeg$, respectively.  These ``expected'' sizes are shown in Figure~\ref{medresults} for reference.  Allowing a wider and wider distribution of possible intrinsic sizes will inflate these ratios, which our simulation predicts.  Figure~\ref{medresults} shows that the median projected size ratios start high and tend toward the predicted value in moving from the Gaussian to the wide to the narrow intrinsic size distribution.  Indeed, if we input a single intrinsic size for all objects, the expected median projected size ratios are always recovered.  In the case where there is no viewing angle dependence (no $\theta_c$), we find a median projected size ratio of 1, as expected---on average, RGs and RLQSOs appear the same size because they have the same average viewing angle.  We note that sample size has virtually no effect on these median values, and that while there is a dependence on the shape of the intrinsic size distribution, the effects are quite small.  The largest difference is due simply to the presence of a distribution of sizes as opposed to a single value.

The minimal effect of the shape of the intrinsic size distribution is also seen in Figure~\ref{meddists}.  In all cases, while there are differences between the results for the narrow (shaded), wide (thick outline), and Gaussian (dotted) histograms, they are small and subtle.  The effects of $\theta_c$ and the sample size are much more prominent.  Increasing $\theta_c$ tends to narrow the projected size ratio distributions.  For a given value of $\theta_c$, a larger sample size will increasingly narrow the distributions (recall that the sample size increases from B89 to B13 to S13).  This has important implications for the use of the median projected size ratio to infer things like a specific value of $\theta_c$, which we will explore further in \S\ref{sec:discuss}.

Table~\ref{simresults} gives the results of simulations for each sample, underlying intrinsic size distribution, and value of $\theta_c$; Figures~\ref{prob1.5results}-\ref{ksresults} show them graphically.  Table~\ref{simresults} is divided into four main sections.  Each of those four sections is divided into two---the top half shows results keeping the original $n_{tot}$ and $n_{RG}/n_{RLQ}$ values, while the bottom gives results with the original $n_{tot}$ but resampling to a 1:1 $n_{RG}/n_{RLQ}$ ratio.  The top three main sections give the probability of various outcomes based on integrating the PDFs.  We discuss each of the four sections in the table in the following subsections.

\subsubsection{Probability of $\langle RG \rangle / \langle RLQ \rangle>$ 1.5}
\begin{figure}
   \includegraphics[width=8.5cm]{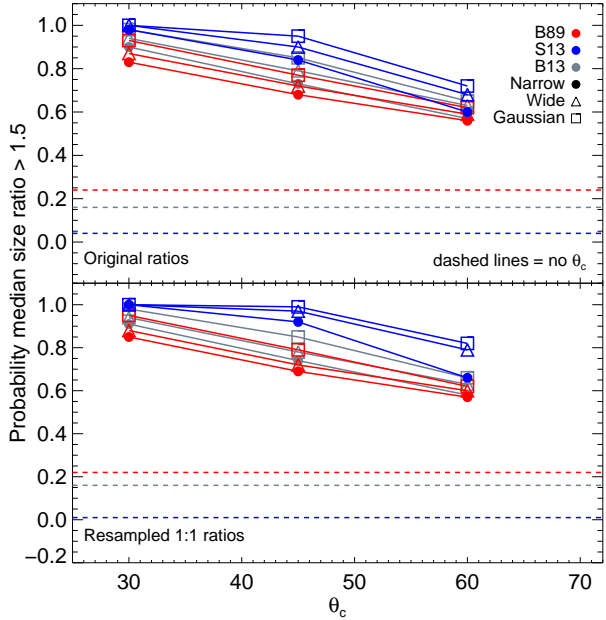}
  \caption{The probability of finding a median size ratio of RGs to QSOs ($\langle {\rm RG} \rangle / \langle {\rm RLQ} \rangle$) greater than 1.5 for each sample (indicated by color), as a function of several parameters; the value of $\theta_c$ (on the x-axis), sample size (increases from B89 to B13 to S13), and the underlying intrinsic size distribution (narrow; filled circles, wide; open triangles, Gaussian; open squares).  The top panel shows results keeping the original $n_{RG}/n_{RLQ}$ ratios for each sample, and the bottom shows the results keeping the same $n_{tot}$ but randomly resampling to an equal number of RGs and RLQSOs.\label{prob1.5results}}
\end{figure}

The first section of Table~\ref{simresults} shows the probability of finding a value of $\langle {\rm RG} \rangle / \langle {\rm RLQ} \rangle$ of 1.5 or greater, and the results are shown graphically in Figure~\ref{prob1.5results}.  The value 1.5 was chosen as a limit because it is the smallest ratio we would expect to find (see Figure~\ref{medresults}), unless the value of $\theta_c$ is significantly larger than 60\arcdeg.  
 
In the case of no orientation unification (no $\theta_c$) and a single intrinsic size, the probability of seeing a projected size difference is essentially zero.  However, one striking result is that when considering the intrinsic size distribution as well as the orientation, depending on the sample size, there is a significant probability that RGs will appear larger than RLQSOs, even if there is no difference in viewing angle to RGs and RLQSOs. This effect has little dependence on the shape of the intrinsic size distribution.  However, for a sample size in the several hundreds, such as S13, the large number of sources overrides the intrinsic size distribution and the probability of finding a projected size difference in the no $\theta_c$ case again approaches zero.  This means that for small sample sizes, as in B89, there is a reasonable probability of finding a projected size difference even if there is no viewing angle difference for RGs and RLQSOs---larger sample sizes are needed to rule out this possibility.

If the hypothesis of unification by orientation is adopted for general values of $\theta_c$, an increase in sample size makes it more likely that a projected size ratio greater than 1.5 will be seen.  However, for an increase in sample size of a factor of nearly 10 (from B89 to S13), the probability only increases by $\sim$20\%.  The effect of the intrinsic size distribution in these cases is not large, but is more significant---wider distributions are more likely to show a projected size difference.  This is reasonable, as very different intrinsic sizes are more likely to be randomly selected if the intrinsic size distribution is broad. Applying different projections---due to the differing allowed viewing angles---to a
survey that samples a broad intrinsic size distribution then makes the projected size ratio even larger.  A ratio of RGs to RLQSOs of $\sim$1 also slightly increases the probability of finding a large projected size ratio, but the effect is not large.

The particular value of $\theta_c$ also affects the probability of finding a projected size difference.  Of course, in reality we do not control $\theta_c$, it just has some intrinsic value (if it exists at all).  However, understanding this dependence is useful as there are ways of estimating $\theta_c$ (e.g.\ B89 estimate it at $\sim$45\arcdeg). Other estimates may be made in the future, and understanding the dependence shown here can be useful in order to determine the robustness of future results.

If we assume the $\theta_c=45\arcdeg$ value from B89 is correct, the results in this analysis suggest that the probability of finding a size ratio that can be interpreted as a significant projected size difference between RGs and RLQSOs in the B89, B13, and S13 experiments ranges from $\sim$65 to 95\%.

\subsubsection{Probability of finding a minimum $\langle RG \rangle / \langle RLQ \rangle$}
\begin{figure}
   \includegraphics[width=8.5cm]{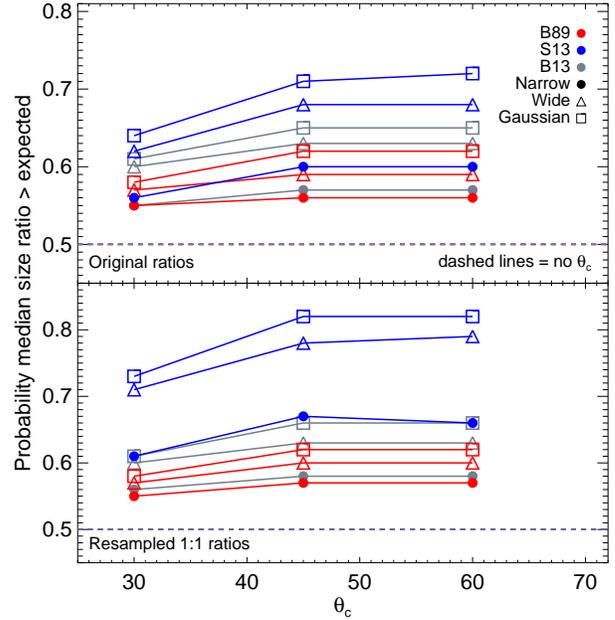}
  \caption{The same as Figure~\ref{prob1.5results}, except showing the probability of finding $\langle RG \rangle / \langle RLQ \rangle$ above the ``expected'' size ratios for a given $\theta_c$ in the idealized, single intrinsic size case.  These median size ratios are 2.5, 1.8, and 1.5 for $\theta_c=30, 40, 60\arcdeg$, respectively, and 1 for the no $\theta_c$ case.  The top panel shows results keeping the original $n_{RG}/n_{RLQ}$ ratios for each sample, and the bottom shows the results keeping the same $n_{tot}$ but randomly resampling to an equal number of RGs and RLQSOs.\label{probcritresults}}
\end{figure}

The second part of Table~\ref{simresults} gives the probabilities of finding a median size ratio larger than the ``expected'' value for each $\theta_c$; 2.5, 1.8, and 1.5 for $\theta_c =30\arcdeg, 45\arcdeg, 60\arcdeg$, respectively.  Essentially, these are the values we would need to obtain, at a minimum, to argue that the assumptions we put into the model are recovered, i.e.\ for example, that we could use the median projected size ratio to find the input $\theta_c$.  The probability of reaching these values is shown in Figure~\ref{probcritresults}.  The probabilities are roughly constant with $\theta_c$ (within $\sim$5\%) for a given sample and intrinsic size distribution. Figure~\ref{meddists} implies why this is the case---as $\theta_c$ decreases, the expected value of the median projected size ratio decreases, and the size ratio distributions shift toward lower values in turn, keeping the probabilities roughly constant.  

Sample size has an even smaller effect on these probabilities than in the previous case, with a factor of $\sim$10 increase in sample size leading to at most a 10\% increase in the probability of detecting at least the expected value.  A ratio of $n_{RG}/n_{RLQ}$ closer to $\sim$1 can add an additional 10\% to the probability that the expected values, or larger, will be recovered.

The results in this analysis indicate that the probability of finding the median projected size ratios (or larger) predicted under unification by orientation in the B89, B13, and S13 experiments ranges from $\sim$55-70\%, again assuming $\theta_c=45\arcdeg$.

\subsubsection{Probability of larger RLQSOs}

\begin{figure}
   \includegraphics[width=8.5cm]{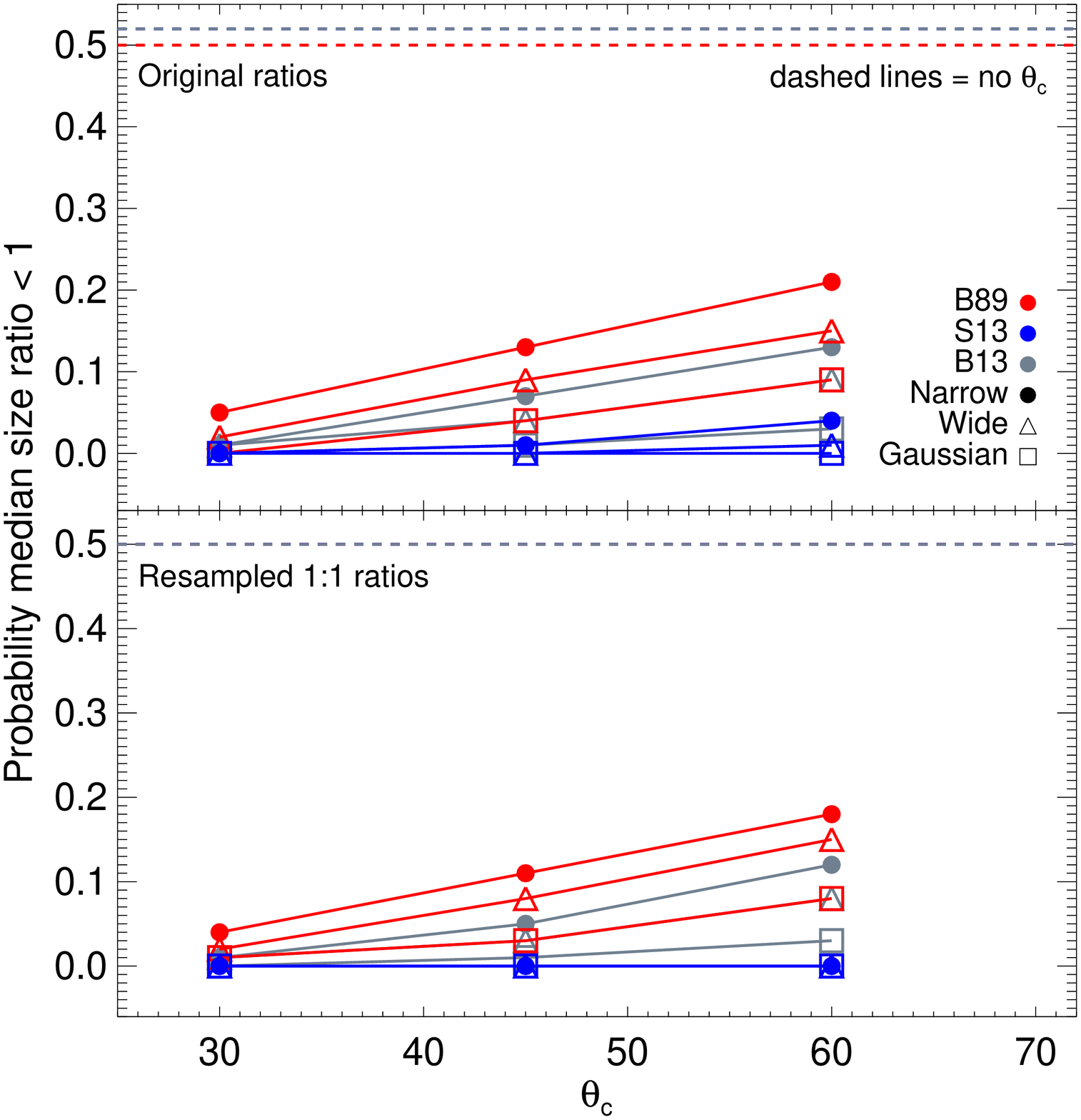}
  \caption{The probability that RLQSOs will appear larger than RGs in projected size ($\langle {\rm RG} \rangle / \langle {\rm RLQ} \rangle<1$.  Panels are the same as Figure~\ref{medresults}.  The top panel shows results keeping the original $n_{RG}/n_{RLQ}$ ratios for each sample, and the bottom shows the results keeping the same $n_{tot}$ but randomly resampling to an equal number of RGs and RLQSOs.\label{fracresults}}
\end{figure}

We can also determine the probability that the S13 and B13 result (that RLQSOs appear larger than RGs) could be recovered if the hypothesis of unification by orientation was correct.  This analysis is shown in the third section of Table~\ref{simresults} and in Figure~\ref{fracresults}, as the probability that a simulation will result in a median projected RG to RLQSO size ratio of less than one.  

We see again that there is a dependence on the value of $\theta_c$, though this becomes less critical as sample sizes increase.  The effect of a wide vs.\ narrow vs.\ Gaussian intrinsic size distribution is also small here, with a change of only a few percent in probability between them.  Resampling the sources to have an equal number ratio also has a minimal effect. 

In this analysis we see that it is possible in the standard orientation picture to randomly find that RLQSOs appear larger than RGs, \textit{but it is extremely unlikely, especially for the large sample size of S13.}  

\subsubsection{Analysis using distribution (KS) tests}
\begin{figure}
   \includegraphics[width=8.5cm]{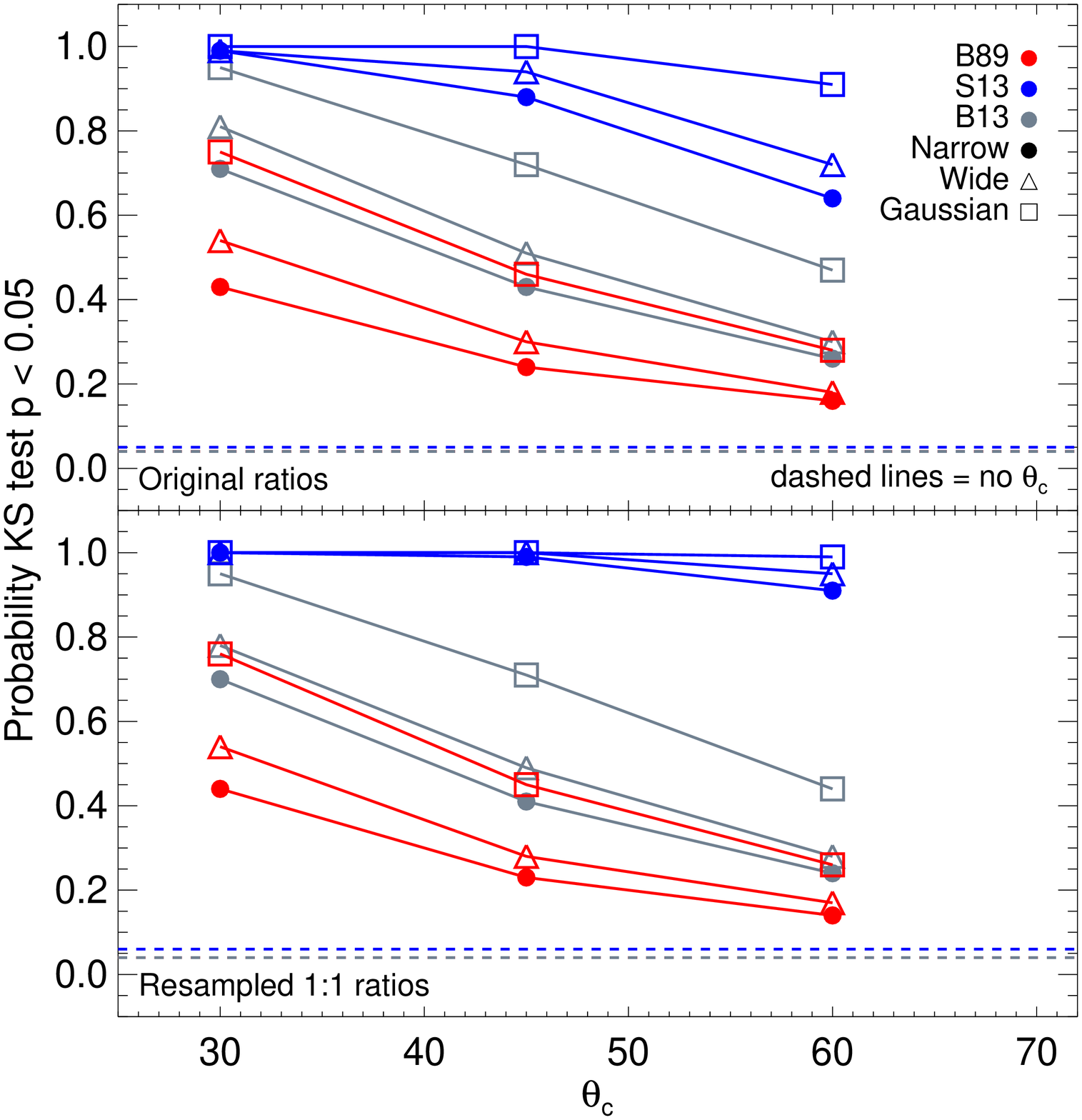}
  \caption{The probability that a KS test will show a significant ($p < 0.05$) difference in projected size distributions of RGs and RLQSOs.  The top panel shows results keeping the original $n_{RG}/n_{RLQ}$ ratios for each sample, and the bottom shows the results keeping the same $n_{tot}$ but randomly resampling to an equal number of RGs and RLQSOs.\label{ksresults}}
\end{figure}

B89, B13 and S13 generally present their results in the context of $\langle {\rm RG} \rangle / \langle {\rm RLQ} \rangle$.  However, when illustrating the projected sizes, they are presented as cumulative distributions.  Whether or not these cumulative distributions are different is quantified by a Kolmogorov-Smirnov (KS) test.  While the authors do not explicitly use the KS test in their analysis, presenting the cumulative distribution functions essentially allows the reader to perform a KS-test-by-eye.  Therefore, we also perform a KS test on the distributions of projected RG and RLQSO sizes in each iteration of our simulations, using a p-value of less than 0.05 to indicate that the projected size distributions appear significantly different.  There are some caveats to the use of this test in assessing significant differences between heterogeneous populations, but, for our highly controlled simulated data, the KS test adequately discriminates different populations.  The results are shown in the bottom section of Table~\ref{simresults} and in Figure~\ref{ksresults}.

The major difference between using a KS test or the median size ratio to identify a projected size difference is that the KS test is much more conservative, in particular for smaller samples.  For the B89 sample, the probability of finding a significant size difference using the KS test is lower by $\sim$40\% for a given $\theta_c$ compared to the probability of finding a median projected size ratio greater than 1.5.  For the larger S13 sample, this difference in probability is much less significant.  As expected, larger samples are more likely to be able to constrain differences in projected size, regardless of the shape of the intrinsic size distribution.  Sample size plays a much larger role in the resulting probability using the KS test than when using the median projected size ratio distributions, with differences in probability of $\sim$80\% from the smallest (B89) to largest (S13) samples.  The ratio of RGs to RLQSOs does not play a very large role in this test, though for the S13 sample, which has the largest difference in $n_{RG}$ and $n_{RLQ}$, we clearly see that having a ratio closer to $\sim$1 increases the chance of seeing a difference in sizes.  It is much less apparent in the other samples, which have a more nearly equal number of RGs and RLQSOs.

Again, the probability of a KS test resulting in $p<0.05$ depends on the value of $\theta_c$.  This probability is more strongly affected by the underlying intrinsic size distributions than in the previous cases, but still behaves as would be expected.  Widening the distribution of intrinsic size is more likely to result in a difference in projected size, as it increases the likelihood that the sizes are different before any projection effects are applied.

Assuming the B89 value of $\theta_c=45\arcdeg$, B89 should have observed a significant projected size difference in 25--30\% of experiments, B13 in 40--50\% of experiments, and S13 in 90--95\% of experiments. Using the Gaussian distribution of intrinsic sizes, these results can increase by up to a further 20\%.

\begin{deluxetable*}{lccccccccccccccccc}
\centering
 \tabletypesize{\footnotesize}
 \tablecaption{Simulation results\label{simresults}}
    \tablehead{
   \colhead{}  &  \colhead{}  &  \colhead{}  &  \colhead{}  &  \multicolumn{4}{c}{Narrow}  & \colhead{} & \multicolumn{4}{c}{Wide}  & \colhead{}  & \multicolumn{4}{c}{Gaussian}     \\       
                                                                                                   \cline{5-8}                                                                 \cline{10-13}                                                        \cline{15-18}                            \\   
   \colhead{Sample} & \colhead{$n_{tot}$} & \colhead{$n_{RG}$} & \colhead{$n_{RLQ}$} & \colhead{no $\theta_c$} & \colhead{$\theta_c = 30\arcdeg$} & \colhead{$45\arcdeg$} & \colhead{$60\arcdeg$} & \colhead{} & \colhead{no $\theta_c$} & \colhead{$\theta_c = 30\arcdeg$} & \colhead{$45\arcdeg$} & \colhead{$60\arcdeg$} & \colhead{}  &   \colhead{no $\theta_c$} & \colhead{$\theta_c = 30\arcdeg$} & \colhead{$45\arcdeg$} & \colhead{$60\arcdeg$}
   }   
   \startdata
  &  &  &  &  \multicolumn{14}{c}{Probability $\langle {\rm RG} \rangle / \langle {\rm RLQ} \rangle >1.5$  (RGs larger)} \\
    \cline{5-18}  \\
B89   &  50    &  33   &  17 & 0.24 & 0.83  & 0.68 & 0.56 & & 0.24 & 0.87  & 0.72 & 0.59 & & 0.16 & 0.93 & 0.77 & 0.62 \\
B13   &   86   &  51   &  35 & 0.17 & 0.90 & 0.73  & 0.57 & & 0.16 & 0.94  & 0.79 & 0.63 & & 0.09 & 0.98 & 0.85 & 0.65 \\
S13   &  466  & 379 &  87 & 0.04 &  0.98 & 0.84 & 0.60 & & 0.04  & 1.00  & 0.90 & 0.68 & & 0.01 & 1.00 & 0.95 & 0.72 \\
\\
\hline  \\
B89   &  50    &  25   &  25    & 0.23 & 0.85 & 0.69 & 0.57 & & 0.22 & 0.88 & 0.72 & 0.60 & & 0.15 & 0.95 & 0.79 & 0.62 \\
B13   &  86    &  43   &  43    & 0.17 & 0.91 & 0.74 & 0.58 & & 0.16 & 0.94 & 0.78 & 0.63 & & 0.08 & 0.98 & 0.85 & 0.66  \\
S13   & 466   & 233 &   233 & 0.01 & 1.00 & 0.92 & 0.66 & & 0.01 & 1.00 & 0.97 & 0.79 & & 0.00 & 1.00 & 0.99 & 0.82  \\
\\
\hline 
\hline \\
  &  &  &  &  \multicolumn{14}{c}{Probability $\langle {\rm RG} \rangle / \langle {\rm RLQ} \rangle >$ expected value (RGs larger)} \\
    \cline{5-18}  \\
B89   &  50    &  33   &  17 & 0.50 & 0.55  & 0.56 & 0.56 & & 0.49 & 0.57  & 0.59 & 0.59 & & 0.50 & 0.58 & 0.62 & 0.62 \\
B13   &   86   &  51   &  35 & 0.51 & 0.55 & 0.57  & 0.57 & & 0.50 & 0.60  & 0.63 & 0.63 & & 0.50 & 0.61 & 0.65 & 0.65 \\
S13   &  466  & 379 &  87 & 0.51 &  0.56 & 0.60 & 0.60 & & 0.50  & 0.62  & 0.68 & 0.68 & & 0.50 & 0.64 & 0.71 & 0.72 \\
\\
\hline  \\
B89   &  50    &  25   &  25    & 0.51 & 0.57 & 0.60 & 0.60 & & 0.50 & 0.55 & 0.57 & 0.57 & & 0.52 & 0.58 & 0.62 & 0.62 \\
B13   &  86    &  43   &  43    & 0.50 & 0.60 & 0.63 & 0.63 & & 0.50 & 0.56 & 0.58 & 0.58 & & 0.50 & 0.61 & 0.66 & 0.66  \\
S13   & 466   & 233 &   233 & 0.51 & 0.71 & 0.78 & 0.79 & & 0.50 & 0.61 & 0.67 & 0.66 & & 0.49 & 0.73 & 0.82 & 0.82  \\
\\
\hline 
\hline \\
  &  &  &  &  \multicolumn{14}{c}{Probability $\langle {\rm RG} \rangle / \langle {\rm RLQ} \rangle <1$ (RLQSOs larger)} \\
    \cline{5-18}  \\
B89   &  50    &  33   &  17 & 0.50 & 0.05  & 0.13 & 0.21 & &  0.50 & 0.02  & 0.09 & 0.15 & & 0.49 & 0.00 & 0.04 & 0.09 \\
B13   &   86   &  51   &  35 & 0.50 & 0.01 & 0.07  & 0.13 &  & 0.50 & 0.01  & 0.04 & 0.09 & & 0.50 & 0.00 & 0.01 & 0.03 \\
S13   &  466  & 379 &  87 & 0.52 & 0.00 & 0.01 & 0.04 &  & 0.52  & 0.00  & 0.00 & 0.01 & & 0.52 & 0.00 & 0.00 & 0.00 \\
\\
\hline  \\
B89   &  50    &  25   &  25    & 0.51 & 0.04 & 0.11 & 0.18 & & 0.50 & 0.02 & 0.08 & 0.15 & & 0.50  & 0.01 & 0.03 & 0.08 \\
B13   &  86    &  43   &  43    & 0.49 & 0.01 & 0.05 & 0.12 & & 0.50 & 0.01 & 0.03 & 0.08 & & 0.50 & 0.00 & 0.01 & 0.03 \\
S13   & 466   & 233 &   233 & 0.50 & 0.00 & 0.00 & 0.00 & & 0.50 & 0.00 & 0.00 & 0.00 & & 0.50 & 0.00 & 0.00 & 0.00 \\
\\
\hline 
\hline \\
  &  &  &  &  \multicolumn{14}{c}{Probability of KS test with $p < 0.05$} \\
  \cline{5-18}  \\
B89   &  50    &  33   &  17 &  0.05 & 0.43 & 0.24 & 0.16 & & 0.04 & 0.54  &  0.30 & 0.18 & & 0.05 & 0.75 & 0.46 & 0.28 \\
B13   &   86   &  51   &  35 & 0.05 & 0.71 & 0.43 & 0.26 & & 0.04  & 0.81  & 0.51 & 0.30 & & 0.05 & 0.95 & 0.72 & 0.47 \\
S13   &  466  & 379 &  87 & 0.05 & 0.99 & 0.88 & 0.64  & & 0.05  & 0.99  & 0.94 & 0.72  & &  0.05 & 1.00 & 1.00 & 0.91 \\
\\
\hline  \\
B89   &  50    &  25   &  25    & 0.03 & 0.44 & 0.23 & 0.14 & & 0.04 & 0.54 & 0.28 & 0.17 & & 0.04  & 0.76 & 0.45 & 0.26  \\
B13   &  86    &  43   &  43    & 0.04 & 0.70 & 0.41 & 0.24 & & 0.04 & 0.78 & 0.49 & 0.28 & & 0.04 & 0.95 & 0.71 & 0.44  \\ 
S13   & 466   & 233 &   233 & 0.05 & 1.00 & 0.99 & 0.91 & & 0.06 & 1.00 & 1.00 & 0.95 & & 0.05 & 1.00 & 1.00 & 0.99  
\enddata
 \tablecomments{The top three sections of the table show simulation results based on integrating the PDFs of median projected sizes from Figure~\ref{meddists} (though normalized to an area of one, not a peak of one as shown in the figure) using various limits.  The top section shows the probability of finding a median projected size ratio greater than 1.5, the second section shows the probability of finding a median projected size ratio greater than the value expected (in the simplest, single intrinsic size scenario) for each value of $\theta_c$, and the third section shows the probability of finding a median projected size ratio less than one (indicating RLQSOs appear larger than RGs, as suggested by the results of B13 and S13).  The last section shows the probability of applying the KS test to the projected size distributions of RGs and RLQSOs and finding a p-value of less than 0.05, indicating a significant probability that the size distributions are different.  Each section is separated into two---the first keeping the original $n_{tot}$ and $n_{RG}/n_{RLQ}$ ratios, and the second showing results when the samples have the same $n_{tot}$ but a 1:1 ratio of $n_{RG}$ to $n_{RLQ}$.  All of these results are shown graphically in Figures~\ref{prob1.5results}-\ref{ksresults}.}
\end{deluxetable*}

\subsection{An ``ideal'' sample size}
\begin{figure}
\centering
   \includegraphics[width=8.5cm]{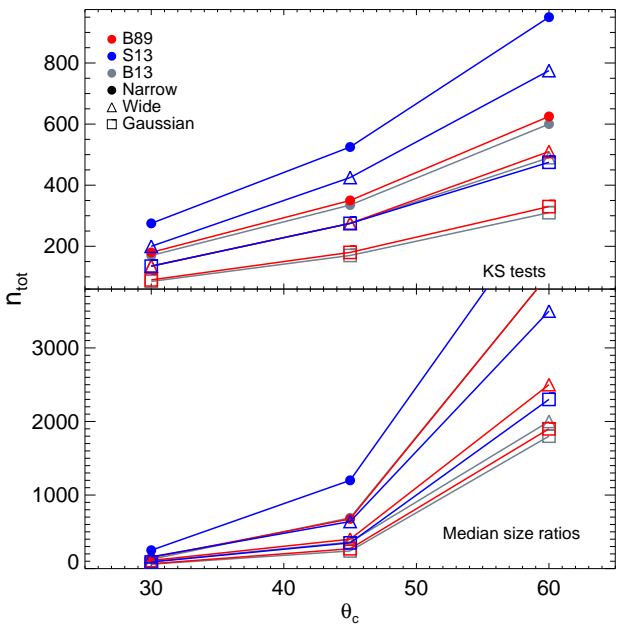}
  \caption{The number of sources needed in order to see a difference in the projected size distributions in $\sim$95\% of experiments, assuming the simple orientation picture is correct.  The top panel shows the number needed if comparisons are made with a KS test, the bottom panel shows the number needed if looking for a median projected size ratio of 1.5 or greater.  For $\theta_c=60\arcdeg$ and the narrow intrinsic size distribution, the required number of sources is $>$4000.  These upper limits are cut off from the figure to better illustrate the other values.\label{ideal}}
\end{figure}

There is another way to approach this problem: under the assumption that the orientation model is correct, what sample size do we really need to reliably see a difference in the projected sizes of RGs and RLQSOs? Our simulations are equipped to answer this question by randomly sampling the original objects to obtain an arbitrary sample size.  We adjust the number of objects needed in each case (i.e.\ for each $\theta_c$ and for each intrinsic size distribution) until the probability of a KS test with $p < 0.05$ is $\sim$95\%, as well as until the probability of finding a median projected size ratio larger than 1.5 is $\sim$95\%.  The results are summarized in Table~\ref{idealsample} and Figure~\ref{ideal}.

The required sample size depends heavily on the value of $\theta_c$, which again we cannot control in reality.  However, we reiterate that this information can be useful if predictions are made about the true value of $\theta_c$.  If we again assume that the canonical B89 value is correct, the B89 and B13 samples are too small to reliably find a projected size difference, particularly in a median size ratio analysis.  The S13 sample is large enough assuming the wider or Gaussian intrinsic size distributions, and nearly large enough for the narrower distribution if tested with a KS test.  For the median projected size comparison, the S13 sample is only large enough with the Gaussian intrinsic size distribution.

Sample sizes must be much larger to detect a projected size difference at a high probability with the median projected size ratio analysis, compared to with a KS test.  In the case of $\theta_c = 60\arcdeg$ the required sample sizes are in the several thousands.  This is because the median projected size ratio distribution is centered near 1.5, and a large sample size is required to sufficiently narrow the distribution such that the high median size tail dominates.  For the original parameters of the samples, S13 requires significantly more objects than B89 or B13.  This is due to the large discrepancy in $n_{RG}$ and $n_{RLQ}$ in the S13 sample.  In fact, if the three samples are simulated to have a 1:1 RG to RLQSO ratio, then the total number of objects needed is the same regardless of the other details of the sample.  

\begin{deluxetable*}{lcccccccccccc}
\centering
 \tabletypesize{\footnotesize}
 \tablecaption{Ideal sample sizes\label{idealsample}}
    \tablehead{
   \colhead{}  &  \colhead{} &  \multicolumn{3}{c}{Narrow}  & \colhead{} & \multicolumn{3}{c}{Wide}  &  &  \multicolumn{3}{c}{Gaussian}     \\       
                                                                    \cline{3-5}                                                                 \cline{7-9}                           \cline{11-13}        \\   
   \colhead{Sample} &  \colhead{$\theta_c$ ($\arcdeg$)} & \colhead{$n_{tot}$} & \colhead{$n_{RG}$}  &  \colhead{$n_{RLQ}$} & & \colhead{$n_{tot}$} & \colhead{$n_{RG}$}  &  \colhead{$n_{RLQ}$} & & \colhead{$n_{tot}$} & \colhead{$n_{RG}$}  &  \colhead{$n_{RLQ}$}
   }   
   \startdata
\multicolumn{13}{c}{KS test}  \\
\cline{1-13} \\
B89  &  30 &    180      &   119       &       61   & &  135     &      92      &     48      &  &    90   & 59    & 31      \\
         &  45 &    350      &    231      &     119   & & 275     &     182     &     93      &  &    180  & 119 & 61      \\
         &  60 &    625      &   413      &     212   &  &  510     &      337    &    173     &  &    330 & 218 & 112     \\ 
\hline \\
B13  &  30 &     170    &     100     &       70    & & 135     &     80     &       55     &  &      85    & 50   & 35    \\
         &  45 &     335    &      198    &      137    & & 275     &    162    &      113   &  &      170 & 100 & 70   \\
         &  60 &     600    &     354     &       246   & & 490     &    289    &      201   &  &      310 & 183 & 127   \\
\hline \\
S13  &  30 &    275   &      223     &       52   & &        200     &      110   &        25    & & 135 & 109 & 26     \\
         &  45 &    525     &     425      &     100   & &     425     &      334   &        81    & & 275 & 223 & 52    \\
         &  60 &    950     &     770      &     180   & &      775     &      628   &        147  & & 475 & 385 & 90    \\
\hline   
\hline \\
\multicolumn{13}{c}{$\langle {\rm RG} \rangle / \langle {\rm RLQ} \rangle >1.5$}  \\
\cline{1-13} \\
B89  &  30 &     150          &     99     &     51      & & 110     &     73       &    34       &  &    66     & 44 & 22        \\
         &  45 &     676          &    446    &   230     & & 400      &    264      &   136      &  &    270 & 158 & 82       \\
         &  60 &     $>4000$ &    2640  &   1360   & & 2500   &     1650     &   850      &  &   1900 & 1254 & 646       \\ 
\hline \\
B13  &  30 &     126          &    74        &       52   & & 96       &    57      &     39       &  &    60     & 35  & 45      \\
         &  45 &     690          &   407       &     283   & & 360     &    212    &   148      &  &    240   & 142 & 98      \\
         &  60 &     $>4000$ &    2360    &   1640  & & 2000  &     1180   &   820      &  &    1800 & 1062 & 738      \\
\hline \\
S13  &  30 &     250          &    203      &       47  & & 160     &   130       &   30       & &     90   & 73  & 17      \\
         &  45 &     1200        &  972        &     228  & & 640    &    518       &    122     & &   350   & 284 &  66   \\
         &  60 &     $>$5000 &   3240     &    760   & & 3500  &    2835    &     665     & &   2300 & 1863 & 437        
\enddata
 \tablecomments{This table shows the sample size needed to obtain a KS test $p < 0.05$ in at least 95\% of the simulations (top) for the properties of each real sample, and as a function of $\theta_c$, for the narrow, wide, and Gaussian distributions.  Also shown are the sample sizes needed to obtain a median projected size ratio of 1.5 or larger with a probability of 95\% (bottom).  The no $\theta_c$ case is omitted as (by definition) it would take an infinitely large sample to see a difference in projected size distributions at such a high probability.}
\end{deluxetable*}

\section{DISCUSSION}
\label{sec:discuss}
In order to test unification by orientation via these kinds of projected size tests, sample selection can clearly play an important role.  It is critical to choose a sample that is unbiased, in particular to orientation.  In a radio flux-limited sample, if selected at too high a frequency, an obvious bias would be toward more beamed (pole-on) sources, which could significantly skew the results, particularly with regard to the RLQSOs if they are really seen more face-on (e.g.\ Urry \& Padovani 1995).  Samples selected over different redshift ranges may also produce conflicting results.  However, we can see in these simulations that the three real samples analyzed behave in exactly the same way with regard to all parameters.  The exception to this is when using the apparent angular size as opposed to the projected linear size, as mentioned in \S 4.1.1.  Using the former, one must be careful to have extremely well-matched (nearly one-to-one) redshifts or the size comparisons can be misleading, even if the distributions of $z$ are statistically similar.  S13 do, in fact, use the apparent angular sizes only, and as we find that the median size ratios can be inflated even by slight mis-matching in $z$, their finding that QSOs are larger than RGs is questionable.  At best, the sizes of the two classes are the same.  Other than this, the biggest differences seen are due simply to the size of the samples, and in some smaller ways the ratio $n_{RG}/n_{RLQ}$.  This seems to indicate that sample selection effects are not a primary reason that the results of B89 and S13/B13 are contradictory, and that the problem really is with the underlying orientation model.

Discussions in B89, S13, and B13 focus on viewing angle, while somewhat ignoring that there is an underlying intrinsic size distribution that might influence projected sizes.  If the median size ratio is used as a proxy for how different the projected sizes are, it is interesting to ask whether the underlying intrinsic size distribution can dilute the role of viewing angle for a given sample size. It appears that the intrinsic size distribution plays a role, but its actual form is relatively unimportant.  Assuming a Gaussian, wide or narrow distribution of intrinsic size, the effect on $\langle {\rm RG} \rangle / \langle {\rm RLQ} \rangle$ is minimal.  Once a distribution is present, the different possible values of $\theta_c$ affect the results more than the shape of the distribution of intrinsic sizes.  

While the shape of the intrinsic size distribution is relatively unimportant to the median projected size tests, the presence of a distribution instead of a single size has important consequences in the interpretation of these kinds of projected size results.  First, it makes it possible to identify a significantly large projected size ratio, even when a viewing angle dependence is not present.  In the small sample of B89, where a difference was indeed found, our simulations show that there is a 10-20\% chance of obtaining this result even if no viewing angle difference between RGs and RLQSOs exists.  Our results also show that the ratio $\langle {\rm RG} \rangle / \langle {\rm RLQ} \rangle$ does not simply scale to a value of $\theta_c$, as the intrinsic size distribution makes this ratio vary quite a bit compared to the single intrinsic size case.  For example, a median projected size ratio of $\sim$2 is not that unlikely to find for almost any sample size and any value of $\theta_c$.  So using this value to obtain (or confirm) $\theta_c=45\arcdeg$, as B89 did, is unconvincing.

A complication in these simulations and in performing this kind of test in reality can arise due to the effects of the so-called ``receding torus'' \citep{Lawrence91}, if such an effect really exists.  This model suggests that as the quasar accretion disk increases in luminosity, it will sublimate away more of the dust in the torus and cause the amount of re-processed IR radiation to decrease as the covering fraction of the dusty material decreases.  This inverse relationship has been observed in many studies \citep[e.g.][]{Arshakian05, Cleary07, Calderone12}, but its explanation assumes that the dust is in a torus-like structure and thus assumes that unification by orientation is (at least partially) true.  If the ``receding torus'' interpretation is correct, then assuming a single $\theta_c$ for all sources is not strictly correct, as each object  has its own value of $\theta_c$ that determines whether it is seen as a RG or RLQSO.  

We could in principle estimate this value for each object in each iteration of our simulations by using an optical luminosity as a proxy for the covering fraction and hence $\theta_c$, which would further restrict the possible $\theta$ randomly assigned.  As a simple test of how this could affect our results, we re-ran the simulations allowing each object in each iteration to be assigned a value of $\theta$ restricted by a randomly generated value of $\theta_c$, uniformly distributed between 20\arcdeg\ and 60\arcdeg\ \citep{Arshakian05}.  In all cases, the results are consistent with what is expected for the average $\theta_c$, in this case 40\arcdeg, and any conclusions would remain the same.  This however assumes that both the RGs and RLQSOs have the same range of $\theta_c$ values, or equivalently (in the ``receding torus'' model) accretion disk luminosities.  Therefore  we also simulate situations where some selection effect in each sample causes the RLQSOs to have more luminous accretion disks than the RGs.  We do this by assigning each RG a random $\theta_c$ between 20\arcdeg\ and 40\arcdeg, and each RLQSO a random $\theta_c$ from 40\arcdeg\ to 60\arcdeg.  In this situation, one is less likely to observe a projected size difference; in terms of the KS test, even with the large sample of S13 a difference is observed in only $\sim$75\% of experiments, which would weaken our conclusions regarding the orientation model.  Thus, we stress that if the ``receding torus'' interpretation is valid, samples built for this projected size test must also be somewhat well-matched in accretion disk luminosity, which may be difficult to constrain for RGs in particular, as by definition we cannot see through to the nuclear emission.

Our modeling also provides the ideal sample size with which to detect a difference in projected size distributions, if one exists due to orientation effects.  This number can be dauntingly large if $\theta_c$ is large ($\gtrsim50\arcdeg$) and especially for samples with few RLQSOs compared to RGs.  The situation is worse for the narrower intrinsic size distribution.  It seems that sample sizes thus far have been too small to identify RGs as much larger than RLQSOs at a high probability.

However, if the assumptions we have made here are valid (and we argue that they are at least reasonable), it seems quite clear that the conflicting results between B89 and S13/B13 \textit{cannot easily be reconciled with a paradigm in which different projected sizes of radio sources result solely from orientation.}  B89, B13, and S13 should have seen a median projected size ratio of over 1.5 in $\sim$75, 80, and 90\% of experiments, respectively.  Only B89 observed such a ratio, and as mentioned above, there is a small but significant probability that B89 could find this ratio by chance even if there is no unification by orientation.  If, instead of a ratio of 1.5, we are looking for a size ratio consistent with a particular value of $\theta_c$, the probability that B89, B13, or S13 would have observed such a ratio is lower.  Because the projected size ratio distributions roughly accumulate around the ``expected'' ratios, huge numbers of sources are needed to recover these values at high probability.  This could make it easier to reconcile conflicting results.

In terms of the KS tests, the B13 sample contains a number of objects for which there is a 50/50 chance of seeing a projected size difference (or closer to 75\% if the Gaussian distribution is correct).  But it seems that B89 would have to have been somewhat fortunate to see this difference given their sample size---it would be apparent in $\sim$25-40\% of experiments, while S13 would simultaneously have to have been very unlucky to miss it---a size difference should be apparent in $>$90\% of samples that comprise the number of objects studied in S13.  

The advantage of using the median projected size ratio comparison over a KS test is that it is much less sensitive to the shape of the real intrinsic size distribution (which is not well constrained) than the KS test.  The KS test however has an advantage in that it requires a smaller overall sample to identify a projected size difference in a large fraction of experiments.

It is possible that B13 or S13 could have failed to measure projected size differences due to random sampling---but finding that RLQSOs appear larger than RGs in the popular paradigm, while also still possible, is very unlikely.  For the S13 sample this result should only occur in $<1$\% of experiments.  So, it appears that the underlying intrinsic size distribution does not matter enough to question the newest results regarding this problem---the S13 and B13 tests have therefore convincingly demonstrated that the simplest schemes that unify RGs and RLQSOs by orientation are insufficient.

\section{CONCLUSIONS \& SUMMARY}
Cast solely in terms of viewing angle, recent results in the literature strongly contradict unification by orientation for radio galaxies and radio-loud quasars, as the prediction that RGs should have larger projected radio-source sizes due to a more edge-on perspective is not confirmed.  However, it is highly unlikely that all sources have the same intrinsic size and that results can be interpreted solely in terms of viewing angle.  We explored the role that the underlying intrinsic size distribution of radio sources plays in this type of analysis, paying particular attention to the question of whether the conflicting results can be reconciled by random sampling of this intrinsic size distribution.  We summarize the results as follows:

\begin{itemize}
\item Whether or not a difference in projected radio-source sizes will be seen in a given sample depends heavily on the critical viewing angle that separates RGs and RLQSOs (if one really exists).  While we cannot control this value in reality, understanding this dependence can be useful for analyzing the robustness of past and future results.

\item Sample size is critical for discriminating differences in projected sizes for RGs and RLQSOs---and must number at the very least in the several hundreds for a robust detection.

\item The chance of detecting a difference in projected size does depend on, but is not particularly sensitive to, the shape of the underlying intrinsic size distribution, at least for the distributions tested here.  The existence of a distribution as opposed to a single intrinsic size is more important, in general, than the form it takes.

\item It appears that sample selection biases are unlikely to be the reason conflicting results have been found, as all three real samples tested here behave in an identical manner for all parameters.  

\item In smaller samples ($\sim$100 or less), there is a significant probability that RGs will appear larger than RLQSOs, \textit{even if there is no difference in viewing angle to the two subsets}, due to random sampling of the intrinsic size distribution.

\item It is highly unlikely to find RLQSOs appearing larger than RGs in the standard orientation model, even with a wide distribution of intrinsic sizes.

\item It is possible, \textit{but very unlikely}, under the simplest unification by orientation model, for conflicting results in the literature to be reconciled.

\end{itemize}

Our simulations demonstrate that there is, indeed, a major problem with using the projected size distribution of luminous radio sources to argue that radio galaxies and radio-loud quasars can be unified by orientation.  Our analyses suggest that if orientation does play a role in our view of radio galaxies and radio-loud quasars, it cannot be the dominant factor based solely on comparison of the apparent sizes of these sources on the sky.

\acknowledgements
We would like to thank Ritaban Chatterjee for helping develop the idea for this project, as well as for useful discussions.  MAD, JCR, and ADM were partially supported by NASA through ADAP award NNX12AE38G and EPSCoR award NNX11AM18A and by the National Science Foundation through grant number 1211112.

This research has made use of the NASA/IPAC Extragalactic Database (NED) which is operated by the Jet Propulsion Laboratory, California Institute of Technology, under contract with the National Aeronautics and Space Administration.

\appendix
B89 report that RGs appear larger than RLQSOs, in contradiction with the results of B13/S13.  As both sets of results rely on measurements of the median of a distribution, and as scientists predominantly work with significances based on {\em the mean}, it is worth reviewing these results in the context of inferring significances from {\em the median}.  To do this, we follow the method outlined in \citet{Gott2001}.  

To locate the 2$\sigma$ limits in the smallest-to-largest ordered list of projected sizes, we first calculate $2 \sigma(r) = 2(4n)^{-1/2}$, where $n$ is either $n_{RG}$ or $n_{RLQ}$.  The variable $r$ ranges from 0 to 1 and has an expected value of 0.5.  $0.5 \pm 2\sigma(r)$ represents the locations (indices) of the 2$\sigma$ size limits in the ordered list of size measurements, normalized by $n$.  Multiplying these normalized indices by $n$ and rounding up to the nearest integer allows us to approximate the location in the list of values of the 2$\sigma$ upper and lower limits.  Reading off these values from the list provides the range of median size values for either RGs or RLQSOs within the 2$\sigma$ errors in each of the above samples.

The 2$\sigma$ range of median sizes for each of the samples are as follows:

\begin{itemize}

\item B13: 75 $< \langle RG \rangle <$ 255 kpc; 30 $< \langle RLQ \rangle <$ 225 kpc

\item B13: 108 $< \langle RG \rangle <$ 188 kpc; 131 $< \langle RLQ \rangle <$ 400 kpc

\item S13: 64 $< \langle RG \rangle <$ 105 kpc; 43 $< \langle RLQ \rangle <$ 164 kpc

\end{itemize}

We see that in all cases there is significant overlap between the median sizes of the subsamples.  Thus, when the errors on the median apparent sizes are taken into account, it is clear that {\em none of these samples are of sufficient size to determine reliably if RGs really appear larger than RLQSOs}, and claims based simply on the median sizes in the samples are highly suspect.

\clearpage


\begin{thebibliography}

\bibitem[Antonucci(1993)]{Antonucci1993} Antonucci, R.  1993, ARA\&A, 31, 473

\bibitem[Antonucci \& Miller(1985)]{Ant85} Antonucci, R.R.J., \& Miller, J.S.\ 1985, \apj, 297, 621 

\bibitem[Arshakian(2005)]{Arshakian05} Arshakian, T.G.  2005, \aa, 436, 817

\bibitem[Barthel(1989)]{Barthel1989}  Barthel, P.D.  1989, \apj, 336, 606

\bibitem[Becker(1995)]{Becker1995} Becker, R. H., White, R. L., \& Helfand, D. J. 1995, \apj, 450, 559

\bibitem[Boroson(1992)]{Bor92} Boroson, T.A.\ 1992, \apjl, 399, L15 

\bibitem[Boroson(2011)]{Boroson2011} Boroson, T.A. 2011, AAS Meeting \#217, \#142.22; Bulletin of the American Astronomical Society, Vol. 43 

\bibitem[Boroson(2013)]{Boroson2013} Boroson, T.A.  2013, in prep

\bibitem[Bridle \& Perley(1984)]{Bridle1984} Bridle, A.H. \& Perley, R.A.  1984, \araa, 22, 319

\bibitem[Calderone et al.(2012)]{Calderone12} Calderone, G., Sbarrato, T., \& Ghisellini, G.  2007, \mnras, 425, L41

\bibitem[Cleary et al.(2007)]{Cleary07} Cleary, K., Lawrence, C.R., Marshall, J.A., Hao, L., \& Meier, D.  2012, \apj, 660, 117

\bibitem[Condon(1998)]{Condon1998} Condon, J., Cotton, W.D., Greissen, E.W., Yin, Q.F., Perley, R.A., Taylor, G.B. \& Broderick, J.J. 1998, \apj, 115, 1693

\bibitem[Elvis(2000)]{Elv00} Elvis, M.\ 2000, \apj, 545, 63 

\bibitem[Fanaroff \& Riley(1974)]{Fanaroff74} Fanaroff, B.L. \& Riley, J.M.  1974, \mnras, 167, 31

\bibitem[Fey et al.(1996)]{Fey96} Fey, A.L., Clegg, A.W., \& Fomalont, E.B.  1996, \apjs, 105, 299

\bibitem[Ghisellini(1993)]{Ghisellini1993} Ghisellini, G., Padovani, P., Celotti, A., \& Maraschi, L.  1993, \apj, 407, 65

\bibitem[Gott et al.(2001)]{Gott2001} Gott, J.R., Vogeley, M.S., Silviu, P., \& Ratra, B.  2001, \apj, 549, 1

\bibitem[Grimes et al.(2005)]{Grimes2005} Grimes, J.A., Rawlins, S., \& Willott, C.J.  2005, \mnras, 359, 1345

\bibitem[Helmboldt et al.(2007)]{Helmboldt07} Helmboldt, J.F., Taylor, G.B., Tremblay, S., Fassnacht, C.D., Walker, R.C., Myers, S.T., Sjouwerman, L.O., Pearson, T.J., Readhead, A.C.S., Weintraub, L., Gehrels, N., Romani, R.W., Healey, S., Michelson, P.F., Blandford, R.D., \& Cotter, G.  2007, \apj, 658, 203

\bibitem[Hickox et al.(2011)]{Hick11} Hickox, R.C., Myers, A.D., Brodwin, M., et al.\ 2011, \apj, 731, 117 

\bibitem[Hine \& Longair(1979)]{Hine1979} Hine, R.G. \& Longair, M.S.  1979, \mnras, 188, 111

\bibitem[Kapahi(1998)]{Kapahi1998} Kapahi, V.K., Athreya, R.M., Van Breugel, W., McCarthy, P.J., Subrahmanya, C.R.  1998, \apjs, 118-275

\bibitem[Kellermann et al.(1998)]{Kellermann98} Kellermann, K.I., Vermeulen, R.C., Zensus, J.A., \& Cohen, M.H.  1998, \aj, 115, 1295

\bibitem[Komatsu(2011)]{Komatsu2011} Komatsu, E., Smith, K.M., Dunkley, J., et al.  2011, \apjs, 192, 18

\bibitem[Laing et al.(1983)]{Laing83} Laing, R.A., Riley, J.M., \& Longair, M.S.\ 1983, \mnras, 204, 151 

\bibitem[Lawrence(1991)]{Lawrence91} Lawrence, A.  1991, \mnras, 252, 586

\bibitem[Leahy \& Perley(1991)]{Leahy91} Leahy, J.P. \& Perley, R.A.  1991, \aj, 102, 537

\bibitem[Lister et al.(2009)]{Lister09} Lister, M.L., Aller, H.D., Aller, M.F., Cohen, M.H., Homan, D.C., Kadler, M., Kellermann, K.I., Kovalev, Y.Y., Ros, E., Savolainen, T., Zensus, J.A., \& Vermeulen, R.C.  2009, \aj, 137, 3718

\bibitem[Mack et al.(1997)]{Mack97} Mack, K.-H., Klein, U., O'Dea, C.P., \& Willis, A.G.  1997, A\&AS, 123, 423

\bibitem[Mullin et al.(2008)]{Mullin08} Mullin, L.M., Riley, J.M., \& Hardcastle, M.J.  2008, \mnras, 390, 595

\bibitem[Netzer(1985)]{Net85} Netzer, H.\ 1985, \mnras, 216, 
63 

\bibitem[Netzer(1987)]{Net87} Netzer, H.\ 1987, \mnras, 225, 
55 

\bibitem[Orr \& Browne(1982)]{Orr82} Orr, M.J.L., \& Browne, I.W.A.\ 1982, \mnras, 200, 1067 

\bibitem[Polatidis et al.(1995)]{Polatidis95} Polatidis, A.J., Wilkinson, P.N., Xu, W., Readhead, A.C.S., Pearson, T.J., Taylor, G.B., \& Vermeulen, R.C.  1995, \apjs, 98, 1

\bibitem[Rengelink(1997)]{Rengelink1997} Rengelink, R. B., Tang, Y., de Bruyn, A. G., Miley, G. K., Bremer, M. N., Roettgering, H. J. A., \& Bremer, M. A. R.  1997, A\&A, 124, 259

\bibitem[Scheuer \& Readhead(1979)]{Sch79} Scheuer, P.A.G., \& Readhead, A.C.S.\ 1979, \nat, 277, 182 

\bibitem[Singal(2013)]{Singal2013} Singal, A.K. \& Singh, R.L.  2013, \apj, 766, 37

\bibitem[Urry \& Padovani(1995)]{Urry1995} Urry, C.M. \& Padovani, P.  1995, \pasp, 107, 803 

\bibitem[Willott et al.(2000)]{Willott2000} Willott, C.J., Rawlings, S., \& Jarvis, M.J.  2000, \mnras, 313, 237

\bibitem[Wylezalek et al.(2013)]{Wylezalek2013} Wylezalek, D., Galametz, A., Stern, D., et al.  2013, \apj, in press

\end{thebibliography}
\end{document}